\documentclass[a4paper,12pt]{article}

\usepackage{amsmath,amsthm,amsbsy,amssymb,amsfonts,graphicx,mathrsfs,bm,accents,cite,xcolor,color}

\makeatletter
\catcode`\@=11
\@addtoreset{equation}{section}

\makeatother

\renewcommand{\thefootnote}{\arabic{footnote}}

\newcommand{\Exp}[1]{\operatorname{e}^{#1}}
\newcommand{\abs}[1]{\lvert{#1} \rvert}
\newcommand{\rmd}{{\mathrm{d}}}
\newcommand{\nn}{\nonumber}
\newcommand{\Lie}{\pounds}
\newcommand{\gLie}{\hat{\pounds}}

\newcommand{\cG}{\mathcal G}\newcommand{\cH}{\mathcal H}
\newcommand{\cK}{\mathcal K}\newcommand{\cL}{\mathcal L}
\newcommand{\cM}{\mathcal M}
\newcommand{\cP}{\mathcal P}

\newcommand{\cS}{\mathcal S}
\newcommand{\cV}{\mathcal V}

\newcommand{\rmT}{\mathrm{T}}

\newcommand{\GL}{\mathrm{GL}}
\newcommand{\OO}{\mathrm{O}}

\newcommand{\ubar}[1]{\,\underline{\phantom{h}}\hskip-6pt {#1}}

\newcommand{\bnabla}{{\mathring{\nabla}}{}}
\newcommand{\bGamma}{{\mathring{\Gamma}}{}}

\newcommand{\bS}{{\mathring{\cS}}{}}
\newcommand{\bs}{{\mathring{s}}{}}
\newcommand{\bX}{{\bm X}}
\newcommand{\bY}{{\bm Y}}

\allowdisplaybreaks[3]

\begin{document}

\begin{titlepage}
\renewcommand{\thefootnote}{\fnsymbol{footnote}}

\begin{flushright}
\parbox{4cm}
{KUNS-2653}
\end{flushright}

\vspace*{1cm}

\begin{center}
{\Large\textbf{
Generalized gravity from modified DFT
}}%
\end{center}
\vspace{1.0cm}

\centerline{
{\large Yuho Sakatani$^{a,b}$},%
\footnote{E-mail address: \texttt{yuho@koto.kpu-m.ac.jp}}
{\large Shozo Uehara$^{a}$},%
\footnote{E-mail address: \texttt{uehara@koto.kpu-m.ac.jp}} 
and {\large Kentaroh Yoshida$^{c}$}%
\footnote{E-mail address: \texttt{kyoshida@gauge.scphys.kyoto-u.ac.jp}}
}

\vspace{0.2cm}

\begin{center}
${}^a${\it Department of Physics, Kyoto Prefectural University of Medicine,}\\
{\it Kyoto 606-0823, \rm Japan}

\vspace*{1mm}

${}^b${\it Fields, Gravity \& Strings, CTPU}\\
{\it Institute for Basic Sciences, Daejeon 34047 \rm Korea}

\vspace*{1mm}

${}^c${\it Department of Physics, Kyoto University,}\\
{\it Kitashirakawa Oiwake-cho, Kyoto 606-8502, \rm Japan}
\end{center}

\vspace*{1mm}

\begin{abstract}

Recently, generalized equations of type IIB supergravity have been derived from the requirement of classical kappa-symmetry of type IIB superstring theory in the Green-Schwarz formulation. These equations are covariant under generalized $T$-duality transformations and hence one may expect a formulation similar to double field theory (DFT). In this paper, we consider a modification of the DFT equations of motion by relaxing a condition for the generalized covariant derivative with an extra generalized vector. In this modified double field theory (mDFT), we show that the flatness condition of the modified generalized Ricci tensor leads to the NS-NS part of the generalized equations of type IIB supergravity. In particular, the extra vector fields appearing in the generalized equations correspond to the extra generalized vector in mDFT. We also discuss duality symmetries and a modification of the string charge in mDFT.

\end{abstract}

\thispagestyle{empty}
\end{titlepage}

\setcounter{footnote}{0}

\tableofcontents

\section{Introduction}

A great progress on the Green-Schwarz string theories has been 
achieved by Tseytlin and Wulff \cite{WT}. In particular, they have shown that 
the requirement of classical kappa-symmetry of type IIB superstring theory 
leads to a set of the generalized equations of type IIB supergravity \cite{AFHRT,WT}, 
rather than the standard ones. 
As an old result, it has been 
well known that the on-shell condition of type II supergravity leads to 
kappa-invariant Green-Schwarz string theories \cite{Townsend}. 
But the converse has not been clarified for long time, more than three decades. 
Actually, it should have been modified by employing the generalized equations of 
type IIB supergravity. 

\medskip

This result indicates that the target spacetime may not be necessarily 
a solution of type IIB supergravity at least at classical level. 
However, if the background is not a solution of 
the standard type IIB supergravity, then the string world-sheet is not Weyl invariant 
but still scale invariant \cite{AFHRT,WT}. 
It has not been clarified 
what is the physical origin or mechanism of breaking the Weyl invariance. 
It is of significance to try to unveil it for discovering the unexplored physics frontier 
hiding behind the generalized supergravity. 

\medskip 

The discovery of the generalized equations was not straightforward. It is worth describing 
that the equations were originally found in the study of 
Yang-Baxter deformations \cite{K,DMV1,MY} of the AdS$_5\times$S$^5$ superstring \cite{DMV2,KMY}. 
In the pioneering work \cite{DMV2}, the standard $q$-deformation \cite{DJ} 
(often called $\eta$-deformation) 
of the AdS$_5\times$S$^5$ superstring was studied and the kappa-invariant 
classical action was constructed. After that, the full background has been determined 
by performing supercoset construction \cite{ABF,ABF2}. 
Remarkably, it was shown that the resulting background does not satisfy the equations 
of motion in type IIB supergravity \cite{ABF2}, while it is associated with 
a solution of type IIB supergravity with a linear dilaton via 
T-dualities along non-isometric directions \cite{HT-sol,HT}. 
Then the generalized equations of type IIB supergravity have been proposed 
so as to involve the background as a solution. At this stage, it seemed likely 
that the generalized equations would heavily depend on the integrable deformation. 
But it is not the case as we have well recognized. The appearance of the generalized equations 
is intrinsic to the kappa-symmetry of the Green-Schwarz formulation, rather than 
a specific integrable deformation. 

\medskip

In addition to the $q$-deformation, a lot of examples of Yang-Baxter deformations 
have been intensively discussed in a series of papers 
\cite{LM-MY,MR-MY,Sch-MY,KMY-SUGRA,MY-duality,
MY-summary,Stijn1,Stijn2,CMY,KY,HvT,ORSY,BW,OvT,Stijn3}. 
By performing supercoset construction \cite{ABF2,KY}, one can derive the full backgrounds. 
Some deformations lead to the standard solutions, but the other ones lead to solutions 
of the generalized type IIB supergravity. The two cases are distinguished by the unimodularity 
condition \cite{BW}. 
For the generalized solutions, see \cite{ABF2,KY,HvT,ORSY,BW}. 
It would be interesting to note that by performing the generalized $T$-duality 
transformations \cite{AFHRT}, generalized solutions may be mapped to solutions of the standard type 
IIB supergravity \cite{HT-sol,ORSY,BW}, some of which are locally equivalent to the undeformed 
AdS$_5\times$S$^5$ \cite{ORSY}. 

\medskip 

An interesting observation is that the generalized equations are covariant under generalized $T$-duality transformations \cite{AFHRT} and hence a manifestly $T$-duality covariant formulation developed in the double field theory (DFT) \cite{Siegel:1993xq,Siegel:1993th,Siegel:1993bj,Hull:2009mi,Hull:2009zb,Hohm:2010jy,Hohm:2010pp,Jeon:2010rw,Hohm:2010xe,Jeon:2011cn,Hohm:2011zr,Hohm:2011dv,Jeon:2011vx,Hohm:2011si,Hohm:2011nu,Jeon:2011sq,Jeon:2012kd,Jeon:2012hp} should be efficient in clarifying the structure of the generalized equations (see \cite{Zwiebach:2011rg,Aldazabal:2013sca,Berman:2013eva,Hohm:2013bwa} for reviews of DFT and its extensions). 
In this paper, we discuss a modification of the equations of motion in DFT which leads to the generalized equations of type IIB supergravity. 
The equations of motion in the conventional DFT is expressed as the generalized Ricci flatness conditions, but in the modified DFT (mDFT), introducing an extra generalized vector, we make a modification in the generalized connection and the equations of motion is expressed as the modified generalized Ricci flatness conditions. 
The mDFT clarifies the $T$-duality symmetry of the generalized equations considerably, and we study the global $\OO(N,N)$ transformations which map a solution of mDFT to other solutions of mDFT. 

\medskip 

We note that the modification to the DFT discussed in this paper is rather mild. 
In section \ref{sec:DFT-generalized}, we consider the modification introducing a null generalized Killing vector $\bX^M$ by hand, 
but as we will discuss in the Addendum to section \ref{sec:conclusion} (which is based on the subsequent paper \cite{Baguet:2016prz} by Baguet, Magro and Samtleben), 
this generalized vector $\bX^M$ can be alternatively generated from the DFT by prescribing a Scherk-Schwarz ansatz \eqref{eq:Scherk-Schwarz} 
and we do not need to go beyond the conventional DFT. 

\medskip 

This paper is organized as follows. 
Section 2 introduces the generalized equations of motion of type IIB supergravity. 
In section 3, we give a concise review of DFT with an emphasis on a relation to the conventional supergravity and the construction of curvature tensors in the doubled spacetime. 
In section 4, we consider a modification of the generalized covariant derivative by introducing an extra generalized vector field. 
We then compute the modified generalized Ricci tensor and derive the generalized equations 
of motion from the modified generalized Ricci flatness conditions. 
We also discuss the $T$-duality transformations which map a solution of the mDFT equations of motion to other solutions, and a modification of the fundamental-string charge in mDFT. 
Finally, we make an attempt to construct the classical action for the mDFT equations of motion. 
Section 5 is devoted to conclusion and outlook.

\section{Generalized equations of type IIB supergravity}

In this section, we shall introduce a set of the generalized equations of type IIB supergravity 
\cite{AFHRT,WT}. The generalized equations have recently been derived from the requirement of 
classical kappa-symmetry of type IIB string theory in the Green-Schwarz formulation \cite{WT}. 
This result indicates that the set of the generalized equations is a weaker form of 
the standard type IIB supergravity and the target spacetime may not be necessarily 
a solution of type IIB supergravity. 
Note also here that the off-shell action that leads to the generalized equations has not been 
constructed yet. Hence this is a generalization only under the equations of motion. 

\medskip 

In fact, a curious thing happens at quantum level 
if the target spacetime does not satisfy the 
usual type IIB supergravity equations of motion. The resulting string world-sheet theory 
is not Weyl invariant but still scale invariant \cite{AFHRT,WT}.%
\footnote{A series of earlier works \cite{Hull:1985rc,Tseytlin:1986tt,Shore:1986hk,Tseytlin:1986ws} discussed a subtle difference between the scale invariance and the Weyl invariance in non-linear sigma models.}
It is still unclear what is the physical origin or mechanism of breaking the Weyl invariance. 
One of our motives in this paper is to tackle this issue. 

\medskip

The generalized equations of motion of type IIB supergravity \cite{AFHRT,WT} are given 
by\footnote{The generalized equations of the fermionic components 
have been determined in \cite{WT}. For simplicity, these equations have not been displayed here.}
\begin{eqnarray}\label{general1}
&&R_{mn}-\frac{1}{4}H_{mkl}H_n{}^{kl}-T_{mn}+D_m X_n + D_n X_m = 0\,, \\ 
&&\frac{1}{2}D^k H_{kmn} + \frac{1}{2} F^k F_{kmn}+\frac{1}{12} F_{mnklp}F^{klp} 
 =X^k H_{kmn}+D_m X_n - D_n X_m\,, \label{general2}\\
&&R-\frac{1}{12}H^2+4D_m X^m - 4X_m X^m = 0\,, \label{general3}\\
&&D^m \mathcal{F}_m - Z^m \mathcal{F}_m - \frac{1}{6}H^{mnk}\mathcal{F}_{mnk}=0\,, 
\qquad I^m \mathcal{F}_m = 0\,, \label{general4} \\
&&D^{k}\mathcal{F}_{kmn}-Z^k \mathcal{F}_{kmn} 
- \frac{1}{6}H^{kpq}\mathcal{F}_{kpqmn}
-(I\wedge \mathcal{F}_1)_{mn}=0\,, \label{general5} \\
&&D^{k}\mathcal{F}_{kmnpq} - Z^k \mathcal{F}_{kmnpq} 
+\frac{1}{36}\epsilon_{mnpqrstuvw}H^{rst}\mathcal{F}^{uvw}
-(I\wedge \mathcal{F}_3)_{mnpq}=0\,. 
\label{general6}
\end{eqnarray}
The 10D spacetime indices are labeled by $m,n,\ldots$\, . 
The first equation (\ref{general1}) is for the metric 
in the string frame $G_{mn}$\,. The matter contribution $T_{mn}$ is given by 
\begin{eqnarray}
T_{mn} \equiv \frac{1}{2}\mathcal{F}_m \mathcal{F}_n 
+\frac{1}{4}\mathcal{F}_{mkl}\mathcal{F}_n{}^{kl}
+\frac{1}{4\times 4!}\mathcal{F}_{mpqrs}\mathcal{F}_n{}^{pqrs}
-\frac{1}{4}G_{mn}\Bigl(\mathcal{F}_k\mathcal{F}^{k}
+\frac{1}{6}\mathcal{F}_{pqr}\mathcal{F}^{pqr}\Bigr)\,.
\end{eqnarray}
Here $\mathcal{F}_m\,,\mathcal{F}_{mnk}\,,\mathcal{F}_{mnkpq}$ are 
the rescaled Ramond-Ramond (R-R) field strengths 
\begin{align}
 \mathcal{F}_{n_1n_2\ldots}=\Exp{\Phi}F_{n_1n_2\ldots}\,,
\end{align}
where $\Phi$ is the dilaton whose motion is described by (\ref{general3})\,. 
The second equation (\ref{general2}) is for the field strength $H_{mnk}$ of the Neveu-Schwarz-Neveu-Schwarz (NS-NS) 2-form $B$-field. 
The fourth, fifth and sixth equations (\ref{general4}), (\ref{general5}) and (\ref{general6}) are 
for the R-R 1-form, 3-form and 5-form field strengths. 

\medskip 

Furthermore, the Bianchi identities for R-R field strengths are also modified as 
\begin{eqnarray}
&&(\rmd\mathcal{F}_1 - Z\wedge \mathcal{F}_1)_{mn} - I^k\mathcal{F}_{mnk}=0\,, \\
&&(\rmd\mathcal{F}_3 - Z\wedge \mathcal{F}_3 + H_3\wedge \mathcal{F}_1)_{mnpq} 
- I^k\mathcal{F}_{mnpqk} = 0\,, \\
&&(\rmd\mathcal{F}_5 - Z\wedge \mathcal{F}_5 + H_3\wedge \mathcal{F}_3)_{mnpqrs} 
+ \frac{1}{6}\epsilon_{mnpqrstuvw}I^t\mathcal{F}^{uvw} = 0\,. 
\end{eqnarray} 

\medskip 

A remarkable point is that equations (\ref{general1})--(\ref{general6}) 
involve three new vector fields $X_m$\,, $I_m$ and $Z_m$\,. Hence the above equations are 
referred to as the generalized equations. 
Actually, two of them are independent because the vector $X_m$ is expressed as 
\begin{eqnarray}
X_m \equiv I_m + Z_m\,.
\end{eqnarray}
Then $I_m$ and $Z_m$ satisfy the following relations: 
\begin{eqnarray}
D_m I_n + D_n I_m = 0\,,\qquad D_m Z_n - D_n Z_m + I^k H_{kmn}=0\,, 
\qquad I^m Z_m = 0\,.\label{IZ}
\end{eqnarray}
The first equation of (\ref{IZ}) is the Killing vector equation.
Assuming that $I_m$ is chosen such that the Lie derivative,
\begin{align}
 \Lie_I B_{mn} = I^k \partial_k B_{mn} + B_{kn}\partial_{m} I^k - B_{km}\partial_n I^k\,,
\end{align}
vanishes, the second equation of (\ref{IZ}) can be solved as
\begin{eqnarray}
Z_m = \partial_m \Phi - B_{mn} I^n\,.
\end{eqnarray}
Thus $Z_m$ can be regarded as a generalization of the dilaton gradient $\partial_m \Phi$\,.
In particular, when $I_m$ vanishes, $Z_m$ becomes $\partial_m \Phi$ and 
the generalized equations (\ref{general1})--(\ref{general6}) are reduced 
to the usual type IIB supergravity equations. 

\medskip 

In the following, we will discuss the embedding of the generalized equations into DFT 
introducing a slight modification.

\section{A brief review of DFT}

The DFT is a manifestly $\OO(D,D)$ $T$-duality covariant formulation of supergravity 
initiated in \cite{Siegel:1993xq,Siegel:1993th,Siegel:1993bj,Hull:2009mi,Hull:2009zb,Hohm:2010jy,Hohm:2010pp}. 
In this section, we shall give a concise review for readers not familiar with DFT. 
This review is mainly focused upon 
introducing the classical action and describing the equations of motion of DFT 
in a geometric language. 

\subsection{The classical action of DFT}

In this subsection, we will introduce the classical action of DFT and 
show that the conventional supergravity action can be reproduced from the DFT action. 

\subsubsection*{Basic ingredients of DFT}

Let us introduce a $2D$-dimensional doubled spacetime which is parameterized by the generalized coordinates defined as
\begin{align}
 x^M \equiv (x^m,\,\tilde{x}_m) \qquad (M=1,\dotsc,2D\,;\ m=1,\dotsc, D) \,. 
\end{align}
Here, $x^m$ are the usual coordinates and $\tilde{x}_m$ are the dual coordinates. 
The doubled spacetime has an $\OO(D,D)$-invariant metric, 
\begin{align}
 \eta \equiv (\eta_{MN}) \equiv \begin{pmatrix} 0 & \delta_m^n \\ \delta^m_n & 0 \end{pmatrix} \,,\qquad 
 (\eta^{MN}) \equiv (\eta^{-1})^{MN} = \begin{pmatrix} 0 & \delta^m_n \\ \delta_m^n & 0 \end{pmatrix} \,,
\end{align}
which is utilized to raise and lower the $2D$-dimensional indices $M,N,\cdots$\,. 
Diffeomorphisms in the doubled spacetime is constrained 
so that this $\OO(D,D)$-structure is preserved. 
In fact, the possible diffeomorphisms, called the generalized diffeomorphisms, are generated 
by the generalized Lie derivative $\gLie_V$ which satisfies 
\begin{align}
 \gLie_V\eta_{MN}=0 \,.
\end{align}
For example, the generalized Lie derivative acts on a generalized vector $W^M(x)$ as
\begin{align}
 \gLie_V W^M = V^K\,\partial_K W^M - \big(\partial_K V^M -\partial^M V_K\big)\, W^K \,. 
\end{align}
Here, $V^M(x)=\bigl(v^m(x),\,\tilde{v}_m(x)\bigr)$ are gauge parameters and $\partial_M \equiv (\partial_m,\,\tilde{\partial}^m)$ are the partial derivatives associated with the generalized coordinates $x^M$. 
In the present formulation of DFT, the consistency requires a condition for all physical fields and gauge parameters. 
This, so-called the strong constraint, can be expressed as
\begin{align}
 \eta^{MN}\,\partial_M A(x)\,\partial_N B(x) = \partial_m A(x)\,\tilde{\partial}^m B(x) + \tilde{\partial}^m A(x)\,\partial_m B(x) = 0 \,,
\end{align}
where $A(x)$ and $B(x)$ are physical fields or gauge parameters.%
\footnote{Formulations of DFT without imposing the strong constraint are investigated, for example, in \cite{Lee:2015qza,Ma:2016vgq}.} 
This condition strongly constrains the coordinate dependence of fields. 
As a result, all fields can, at most, depend on half of the doubled coordinates. 
For example, the strong constraint can be satisfied if fields do not depend 
on the dual coordinates, i.e., $\tilde{\partial}^m A(x)=\tilde{\partial}^m B(x)=0$\,. 

\medskip 

In addition to the $\OO(D,D)$-invariant metric, the doubled spacetime also has 
a dynamical metric $\cH_{MN}(x)$, which is called the generalized metric. 
This metric is defined to admit the following decomposition:
\begin{align}
 \cH_{MN} = (\cV^\rmT\,\cV)_{MN} \,,\qquad 
 \cV \in \OO(D,D) \,.
\end{align}
Namely, it parameterizes the coset $\OO(D,D)/\left(\OO(D)\times \OO(D)\right)$ 
and satisfies
\begin{align}
 (\cH^{-1})^{MN} = \cH^{MN} \equiv \eta^{MK}\,\cH_{KL}\,\eta^{LN} \,, \qquad
 \det(\cH_{MN})=1 \,.
\end{align}

\medskip 

Let us consider here the volume form on a doubled spacetime. 
The role of it is played by $\Exp{-2d(x)}$\,, 
where $d(x)$ is the $T$-duality invariant dilaton (sometimes called the DFT dilaton). 
Indeed, under a generalized diffeomorphism, it is defined to behave as
\begin{align}
 \delta_V \Exp{-2d(x)} = \gLie_V \Exp{-2d(x)} \equiv \partial_M \bigl(\Exp{-2d(x)}V^M\bigr) \,. 
\end{align}
This is quite similar to the behavior of the volume form in general relativity, 
\begin{align}
 \Lie_v \sqrt{\abs{G}} = \partial_m \bigl(\sqrt{\abs{G}}\,v^m\bigr)\,.
\end{align}

\subsubsection*{The classical action of DFT}

From the above setups, we can write down the classical action of DFT, which describes the dynamics of the generalized metric $\cH_{MN}(x)$ 
and the $T$-duality invariant dilaton $d(x)$:
\begin{align}
\begin{split}
 S &= \int\! \rmd^{2D}x\, \cL \,,
\\
 \cL &\equiv \Exp{-2d}\Bigl(\,\frac{1}{8}\,\cH^{MN}\,\partial_M \cH^{KL}\partial_N \cH_{KL} -\frac{1}{2}\,\cH^{KL}\,\partial_L\cH^{MN}\,\partial_N \cH_{KM} 
 + 4\partial_M d\,\partial_N\cH^{MN} 
\\
 &\qquad\quad\ -4\,\cH^{MN}\,\partial_M d\,\partial_N d
 - \partial_M\partial_N\cH^{MN} 
 +4\,\cH^{MN}\,\partial_M\partial_N d\,\Bigr) \,. 
\end{split}
\label{eq:DFT-action}
\end{align} 
This action is manifestly invariant under a global $\OO(D,D)$ symmetry which rotates the $2D$-dimensional indices (see section \ref{sec:DFT-mDFT} for more details). 
As long as the strong constraint is imposed, it is also invariant under a local symmetry generated by the generalized Lie derivative \cite{Hohm:2010pp} although the invariance is not manifest in the above expression. 

\subsubsection*{A relation to the conventional supergravity}

In order to reproduce the conventional supergravity action, it is first necessary to remove 
the dual-coordinate dependence from $\cH_{MN}(x)$ and $d(x)$, 
which restricts the partial derivatives in \eqref{eq:DFT-action} into the form 
$\partial_M=(\partial_m,\,0)$\,. 
Next, the following parameterizations for $\cH_{MN}(x)$ and $d(x)$ are supposed 
in terms of the conventional massless fields in the NS-NS sector, 
$\{G_{mn}(x),\,B_{mn}(x),\,\Phi(x)\}$: 
\begin{align}
 \cH_{MN} = \begin{pmatrix} G_{mn}-B_{mp}\,G^{pq}\,B_{qn} & B_{mk}\,G^{kn} \\
  -G^{mk}\,B_{kn} & G^{mn}
 \end{pmatrix}\,, \qquad 
\Exp{-2d}= \Exp{-2\Phi}\sqrt{\abs{G}}\,. 
\label{eq:cH-parametrization}
\end{align}
Then, from a straightforward calculation, the classical action \eqref{eq:DFT-action} 
can be recast into the conventional supergravity action describing 
the NS-NS sector fields (up to a boundary term):
\begin{align}
 S= \int\!\rmd^Dx\Exp{-2\Phi}\sqrt{\abs{G}}\,\left[ 
 R + 4\,\abs{\rmd \Phi}^2 - \frac{1}{2}\,\abs{H_3}^2\right] \,.
\end{align}
Here, $D_m$, $R_{mn}$, and $R$ are the conventional covariant derivative and 
Ricci tensors associated with the metric $G_{mn}$. 
The metric $G_{mn}$ is used to raise and lower the $D$-dimensional indices $m,n,\cdots$\,. 
The following convention is employed for the curvature tensors, 
\begin{align*}
\begin{split}
 D_m v^n &\equiv \partial_m v^n + \gamma_{mk}^{\ n}\,v^k \,,\qquad 
 \gamma_{mn}^{\ k} \equiv \frac{1}{2}\,G^{kl}\,\bigl(\partial_m G_{nl}
 +\partial_n G_{ml}-\partial_l G_{mn}\bigr) \,, \\
 R^p{}_{qmn} &\equiv \partial_m \gamma_{nq}^{\ p} -\partial_n \gamma_{mq}^{\ p} 
  +\gamma_{mr}^{\ p} \,\gamma_{nq}^{\ r} - \gamma_{nr}^{\ p}\,\gamma_{mq}^{\ r} \,,
  \qquad 
 R_{mn}\equiv R^k{}_{mkn} \,,\qquad R\equiv G^{mn}\,R_{mn} \,.
\end{split}
\end{align*}
Note that, by choosing the so-called canonical section $\partial_M=(\partial_m,\,0)$, which satisfies the strong constraint, 
and also choosing the parameterizations \eqref{eq:cH-parametrization}, a local symmetry, generated 
by the generalized Lie derivative $\gLie_V$ with gauge parameters $V^M=(v^m,\,\tilde{v}_m)$\,, 
is reduced to a local symmetry of the conventional supergravity generated 
by the conventional diffeomorphism and the $B$-field gauge transformation,
\begin{align}
 \delta G_{mn} = \Lie_v G_{mn}\,,\qquad 
 \delta B_{mn} = \Lie_v B_{mn} + \partial_m \tilde{v}_n - \partial_n \tilde{v}_m\,.
\end{align}

\medskip 

It should be remarked that the DFT action can be extended to support the R-R fields \cite{Hohm:2011zr,Hohm:2011dv,Jeon:2012kd} and fermion fields \cite{Jeon:2011vx}, and the supersymmetric action is constructed in \cite{Hohm:2011nu,Jeon:2011sq,Jeon:2012hp}. 

\medskip 

In the following, we will concentrate only on the NS-NS sector. 
A generalization to include the R-R fields and fermions is left as a future problem. 

\subsection{Generalized connection and generalized Ricci tensor}
\label{sec:generalized-conection}

In this subsection, we will introduce the generalized covariant derivative 
and the generalized Ricci tensors \cite{Siegel:1993th,Jeon:2010rw,Jeon:2011cn,Hohm:2011si}. 
The generalized Ricci tensors are useful in manifesting the covariance or invariance of various quantities 
under the generalized diffeomorphisms. For example, although the local gauge invariance 
of the DFT action was not manifest from \eqref{eq:DFT-action}, the invariance becomes manifest 
because the Lagrangian density $\cL$ can be identified with a product of the volume factor 
and the generalized Ricci scalar. 
This is a generalization of the Einstein-Hilbert action for the general gravity. 
The equations of motion of DFT can also be expressed 
as the generalized Ricci flatness conditions and the covariance becomes manifest. 

\subsubsection*{Generalized covariant derivative}

Let us define the generalized covariant derivative as%
\footnote{We will basically follow the convention of \cite{Jeon:2010rw,Jeon:2011cn}.}
\begin{align}
 \nabla_M V^N \equiv \partial_M V^N + \Gamma_M{}^N{}_K\,V^K\,,\qquad 
 \nabla_M W_N \equiv \partial_M W_N - \Gamma_M{}^K{}_N\,W_K\,. 
\end{align}
In the conventional formulation, the following four conditions are assumed
\cite{Jeon:2010rw,Jeon:2011cn,Hohm:2011si}:
\begin{itemize}
 \item[(1)] The compatibility with the $\OO(D,D)$ invariant metric: 
\begin{align}
 \nabla_K \eta_{MN}=0\,.
\end{align}
This condition is equivalent to the anti-symmetricity in the last two indices, 
\begin{align}
 \Gamma_{MPQ} = \Gamma_{M[PQ]}\,.
\end{align}

\item[(2)] For arbitrary generalized tensors, the following condition is imposed: 
\begin{align}
 \gLie_V=\gLie^{\nabla}_V\,.
\end{align}
Here $\gLie_V^{\nabla}$ denotes a generalized Lie derivative 
with a generalized covariant derivative, instead of a partial derivative. 
That is, its action on a generalized vector $W^M$ takes the form,
\begin{align}
 \gLie^{\nabla}_V W^M \equiv V^N\,\nabla_N W^M - 
 \bigl(\nabla_N V^M- \nabla^M V_N \bigr)\,W^N\,.
\end{align}
By using the condition (1), i.e., $\Gamma_{MNK} =\Gamma_{M[NK]}$, 
the condition (2) can be recast into \cite{Jeon:2010rw} 
\begin{align}
 \Gamma_{[MNK]} = 0 \,. 
\end{align}
This condition is interpreted as the torsion-free condition \cite{Hohm:2011si} 
because the difference $\gLie_V-\gLie^{\nabla}_V$ is regarded as the generalized torsion tensor. 

\item[(3)] 
The generalized metric is required to be covariantly constant:
\begin{align}
 \nabla_K \cH_{MN}=0\,.
\end{align}
For our later discussion, it is helpful to introduce here the projectors defined as 
\begin{align}
\begin{split}
 &P_{MN} \equiv \frac{1}{2}\, \bigl(\eta_{MN} + \cH_{MN}\bigr)\,,\qquad 
 \bar{P}_{MN} \equiv \frac{1}{2}\, \bigl(\eta_{MN} - \cH_{MN}\bigr) \,,
\\
 &P_M{}^K\,P_K{}^N = P_M{}^N \,, \qquad 
 \bar{P}_M{}^K\,\bar{P}_K{}^N = \bar{P}_M{}^N \,, \qquad
 P_M{}{}^N + \bar{P}_M{}^N = \delta_M{}^N\,. 
\end{split}
\end{align} 
From the conditions (1) and (3), it is easy to see that the projections are covariantly constant,
\begin{align}
 \nabla_M P_K{}^L = 0 \,, \qquad \nabla_M \bar{P}_K{}^L = 0\,.
\end{align}

\item[(4)] The dilaton $d(x)$ is also required to be covariantly constant: 
\begin{align}
 \nabla_M d=0\,. 
\end{align}
The factor $\Exp{-2d}$ behaves as a scalar density with a unit weight, 
hence this condition can be written as
\begin{align}
 \nabla_M \Exp{-2d} = \partial_M \Exp{-2d} + \Gamma_K{}^K{}_M\Exp{-2d} = 0 \,,
\end{align}
or equivalently,
\begin{align}
 \nabla_M d = \partial_M d + \frac{1}{2}\,\Gamma_K{}^K{}_M = 0 \,.
\label{eq:dilaton-cov-const}
\end{align}
\end{itemize}

\subsubsection*{Explicit form of the connection}

The four conditions (1)--(4) can mostly determine the explicit form of $\Gamma_{MNK}$ 
in terms of $\cH_{MN}(x)$ and $d(x)$\,. 
We shall explain here the outline of determining it 
by following \cite{Jeon:2010rw,Jeon:2011cn,Hohm:2011si}. 
It is convenient to employ the projections \cite{Hohm:2011si}:%
\footnote{Note that our projectors $(P,\,\bar{P})$ are the same as the ones in 
\cite{Jeon:2010rw,Jeon:2011cn} and correspond to $(\bar{P},\,P)$ in \cite{Hohm:2011si}.}
\begin{align}
 W_{\ubar{M}} \equiv P_M{}^N \, W_N\,, \qquad 
 W_{\bar{M}} \equiv \bar{P}_M{}^N \, W_N \,,\qquad 
 W_M = W_{\ubar{M}} + W_{\bar{M}} \,.
\end{align}
From the conditions (1) and (3), this projection is consistent 
with the generalized covariant derivative 
(i.e.,~the projection commutes with the generalized covariant derivative each other). 
The condition (1) also ensure that the barred/under-barred indices can be raised or lowered consistently 
(i.e.,~the projection commutes with the raising/lowering operations one another). 
Then, for example, the following relations are satisfied:
\begin{align}
 W^{\ubar{M}}\, Y_{\bar{M}} = 0\,,\qquad 
 W^M\, Y_M = W^{\ubar{M}}\, Y_{\ubar{M}} + W^{\bar{M}}\, Y_{\bar{M}} \,.
\end{align}
These will be useful in our later discussions. 

\medskip 

The properties $\Gamma_{MNK} =\Gamma_{M[NK]}$ and $\Gamma_{[MNK]}=0$ 
allow us to expand $\Gamma_{MNK}$ as
\begin{align}
 \Gamma_{MNK} 
 &= \Gamma_{\ubar{M}\ubar{N}\ubar{K}}{} + \Gamma_{\bar{M}\bar{N}\bar{K}}{} 
   + \bigl(\Gamma_{\ubar{M}\ubar{N}\bar{K}} 
        - \Gamma_{\ubar{M}\ubar{K}\bar{N}} 
        - \Gamma_{\ubar{N}\ubar{K}\bar{M}} 
        + \Gamma_{\ubar{K}\ubar{N}\bar{M}} \bigr)
\nn\\
 &\quad + \bigl(\Gamma_{\bar{M}\bar{N}\ubar{K}} 
        - \Gamma_{\bar{M}\bar{K}\ubar{N}} 
        - \Gamma_{\bar{N}\bar{K}\ubar{M}} 
        + \Gamma_{\bar{K}\bar{N}\ubar{M}} \bigr) \,.
\end{align}
Thus, all we have to do is to determine the following components \cite{Hohm:2011si}:
\begin{align}
 \Gamma_{\ubar{M}\ubar{N}\ubar{K}} \,,\qquad \Gamma_{\bar{M}\bar{N}\bar{K}} \,,\qquad
 \Gamma_{\ubar{M}\ubar{N}\bar{K}} \,,\qquad \Gamma_{\bar{M}\bar{N}\ubar{K}} \,.
\end{align}
The last two are determined only from the conditions (1) and (3) 
\cite{Jeon:2010rw,Jeon:2011cn,Hohm:2011si};
\begin{align}
\begin{split}
 \Gamma_{\ubar{M}\ubar{N}\bar{K}} = -(P\,\partial_{\ubar{M}}P\,\bar{P})_{NK} \,,\qquad
 \Gamma_{\bar{M}\bar{N}\ubar{K}} = (\bar{P}\,\partial_{\bar{M}}P\,P)_{NK} \,. 
\end{split}
\end{align}
On the other hand, $\Gamma_{\ubar{M}\ubar{N}\ubar{K}}$ and 
$\Gamma_{\bar{M}\bar{N}\bar{K}}$, depend on the contracted components 
$\Gamma_M\equiv \Gamma_K{}^K{}_M$\,, which are determined 
from \eqref{eq:dilaton-cov-const}. 
In fact, the following expressions can be obtained:
\begin{align}
\begin{split}
 \Gamma_{\ubar{M}\ubar{N}\ubar{K}} 
 &= \frac{2}{D-1}\, P_{M[N}\, P_{K]}{}^L\, \left(
\Gamma_L 
 - 2\,(P\,\partial^QP\,\bar{P})_{[QL]}\right) 
+ \tilde{\Gamma}_{\ubar{M}\ubar{N}\ubar{K}} \,,
\\
 \Gamma_{\bar{M}\bar{N}\bar{K}} 
 &= \frac{2}{D-1}\, \bar{P}_{M[N}\, \bar{P}_{K]}{}^L\, \left(\Gamma_L 
 - 2\,(P\,\partial^QP\,\bar{P})_{[QL]}\right) + \tilde{\Gamma}_{\bar{M}\bar{N}\bar{K}} \,.
\end{split}
\end{align}
Here, $\tilde{\Gamma}_{\bar{M}\bar{N}\bar{K}}$ are undetermined (or unphysical) 
quantities satisfying
\begin{align}
 \eta^{MK} \,\tilde{\Gamma}_{\ubar{M}\ubar{N}\ubar{K}} = 0 \,, \qquad
 \eta^{MK} \,\tilde{\Gamma}_{\bar{M}\bar{N}\bar{K}} = 0 \,.
\end{align}
They should be projected out from any physical expressions, such as the action and the equations of motion. 

\medskip

By gathering the results obtained so far, the generalized connection is given by 
\begin{align}
\begin{split}
 \Gamma_{MNK} &= \widehat{\Gamma}_{MNK} + \Sigma_{MNK}\,,\qquad 
 \Sigma_{MNK} \equiv \tilde{\Gamma}_{\ubar{M}\ubar{N}\ubar{K}} 
 + \tilde{\Gamma}_{\bar{M}\bar{N}\bar{K}}\,,
\\
 \widehat{\Gamma}_{MNK} &= 2(P\,\partial_M P\,\bar{P})_{[NK]}
    -2\,\big(P_{[N}{}^P\,P_{K]}{}^{Q} - \bar{P}_{[N}{}^P\,\bar{P}_{K]}{}^Q\big)\,\partial_P P_{QM}
\\
  &\quad +\frac{2}{D-1}\,\big(P_{M[N}\,P_{K]}{}^L
                        +\bar{P}_{M[N}\,\bar{P}_{K]}{}^L\big)\,
   \big(\Gamma_L -2\,(P\,\partial^Q P\,\bar{P})_{[QL]}\big) \,.
\end{split}
\label{eq:connection-components}
\end{align}
The contracted components $\Gamma_M$ can be expressed as 
\begin{align}
 \Gamma_M = -2\, \partial_M d \,, 
\end{align}
due to the condition (4), but note that the result \eqref{eq:connection-components} 
itself is independent of the explicit form of $\Gamma_M$\,. 
This observation will play a crucial role in section \ref{sec:DFT-generalized}.

\subsubsection*{Generalized Ricci tensors}

With generalized connections, a generalized Riemann tensor $S_{MNKL}$
can be defined as \cite{Jeon:2010rw,Jeon:2011cn} 
\begin{align}
 S_{MNKL} \equiv \frac{1}{2}\,\bigl(R_{MNKL} + R_{KLMN} 
 - \Gamma_{PMN} \,\Gamma^P{}_{KL} \bigr) \,. 
\end{align}
Here, $R_{PQMN}=\eta_{PR}\, R^R{}_{QMN}$ is a non-tensorial curvature defined as 
\begin{align}
 R^P{}_{QMN} \equiv \partial_M \Gamma_N{}^P{}_Q -\partial_N \Gamma_M{}^P{}_Q 
 +\Gamma_M{}^P{}_R\,\Gamma_N{}^R{}_Q - \Gamma_N{}^P{}_R\,\Gamma_M{}^R{}_Q\,. 
\end{align}
The conditions (1) and (2) enable us to show the following symmetries 
(which are satisfied by the conventional Riemann tensor) \cite{Jeon:2010rw}:
\begin{align}
 S_{MNKL} = S_{[MN]KL} = S_{KLMN} \,,\qquad 
 S_{MNKL} = S_{MN[KL]} \,, \qquad 
 S_{[MNK]L} = 0 \,.
\end{align}
By using the generalized Riemann tensor $S_{MNKL}$, one can define the generalized Ricci tensor%
\footnote{The generalized Ricci tensor here is related to the conventional definition \cite{Jeon:2010rw,Jeon:2011cn,Hohm:2011si} in the following manner: $\cS_{MN}\rvert_{\text{here}}=-2\,\cS_{MN}\rvert_{\text{conventional}}$\,.}
$\cS_{MN}$ and the generalized Ricci scalar $\cS$ as
\begin{align}
 \cS_{MN} &\equiv -2\,\bigl(P_M{}^K\,\bar{P}_N{}^L+\bar{P}_M{}^K\,P_N{}^L \bigr)\, S^P{}_{KPL} \,,
\\
 \cS &\equiv \bigl(P^{MK}\,P^{NL}-\bar{P}^{MK}\,\bar{P}^{NL}\bigr)\, S_{MNKL} \,.
\end{align}
These quantities are constructed such that the contribution from the unphysical connection $\Sigma_{MNK}$ is completely removed \cite{Hohm:2011si}. 
In other words, they are fully covariant quantities in the semi-covariant formulation \cite{Jeon:2010rw,Jeon:2011cn}.

\medskip 

By using the explicit form of the generalized connection \eqref{eq:connection-components} and 
$\Gamma_M = -2\,\partial_M d$, the generalized Ricci tensors, $\cS_{MN}$ and $\cS$, can be expressed in terms of $\cH_{MN}(x)$ and $d(x)$ \cite{Hohm:2010jy};
\begin{align}
 \cS_{MN}&= -2\,\bigl(P_M{}^K\,\bar{P}_N{}^L + \bar{P}_M{}^K\,P_N{}^L\bigr)\, \cK_{KL} \,,
\\
 \cK_{MN} &\equiv \frac{1}{8}\,\partial_M \cH^{KL}\,\partial_N \cH_{KL} 
 - \frac{1}{2}\,\partial_{(M|} \cH^{KL}\,\partial_K \cH_{|N)L}
 + 2\,\partial_M \partial_N d 
\nn\\
 &\quad + \bigl(\partial_K - 2\,\partial_K d\bigr)\,\Bigl( 
        \frac{1}{2}\,\cH^{KL}\,\partial_{(M|} \cH_{|N)L} 
        + \frac{1}{2}\,\cH^L{}_{(M|}\,\partial_L \cH^K{}_{|N)} 
        -\frac{1}{4}\,\cH^{KL}\,\partial_L \cH_{MN} \Bigr) \,,
\\
 \cS&= \frac{1}{8}\,\cH^{MN}\,\partial_M \cH^{KL}\partial_N \cH_{KL} -\frac{1}{2}\,
 \cH^{KL}\,\partial_L\cH^{MN}\,\partial_N \cH_{KM} 
 + 4\partial_M d\,\partial_N\cH^{MN} 
\nn\\
 &\quad -4\,\cH^{MN}\,\partial_M d\,\partial_N d
 - \partial_M\partial_N\cH^{MN} 
 +4\,\cH^{MN}\,\partial_M\partial_N d \,.
\end{align}

\medskip

Let us recall here that the classical DFT action \eqref{eq:DFT-action} is given by 
\begin{align}
 S = \int\! \rmd^{2D}x\, \Exp{-2d} \cS \,.
\end{align}
The volume form $\Exp{-2d}$ and the generalized Ricci scalar $\cS$ behaves 
as a scalar density and a scalar, respectively, under generalized diffeomorphisms. 
Hence, under an infinitesimal generalized diffeomorphism along a generalized vector $V^M$, 
the action transforms as
\begin{align}
 \delta_V S = \int\! \rmd^{2D}x\, \partial_M\bigl(\Exp{-2d} \cS\,V^M\bigr)\,.
\end{align}
Thus, the DFT action is manifestly invariant under the generalized diffeomorphisms, 
at least in the absence of the boundary. 

\medskip 

The equations of motion of DFT can be straightforwardly obtained from the action 
\eqref{eq:DFT-action}, and the result can be summarized as
\begin{align}
 \cS_{MN} =0 \,,\qquad \cS = 0\,. 
\end{align}
It is also possible to define the generalized Einstein tensor as \cite{Park:2015bza}
\begin{align}
 \cG_{MN} \equiv \cS_{MN} - \tfrac{1}{2}\,\cH_{MN}\, \cS \,. 
\end{align}
This tensor satisfies
\begin{align}
 \cS_{MN} = \bigl(P_M{}^K\,\bar{P}_N{}^L+\bar{P}_M{}^K\,P_N{}^L \bigr)\,\cG_{KL}\,,\qquad 
 \cS = -\tfrac{2}{D}\,P^{MN}\,\cG_{MN} 
 = \tfrac{2}{D}\,\bar{P}^{MN}\,\cG_{MN} \,,
\end{align}
and the differential Bianchi identity \cite{Siegel:1993th,Kwak:2010ew,Hohm:2010xe}
\begin{align}
 \nabla_M \cG^{MN} = 0\,. 
\label{eq:diff-Bianch}
\end{align}
By using the generalized Einstein tensor, the equations of motion is written in a simple form:
\begin{align}
 \cG_{MN} = 0\,.
\end{align}
The covariance of the equations of motion is manifest since the generalized Einstein tensor transforms as a generalized tensor under the global $\OO(D,D)$ transformations as well as the local generalized diffeomorphisms.

\subsubsection*{Equations of motion in the conventional formulation}

For later convenience, let us describe the relation between $\cG_{MN}=0$ and 
the equations of motion in the conventional supergravity. 

\medskip 

Note here that $\cS_{MN}=\cS_{\ubar{M}\bar{N}}+\cS_{\bar{M}\ubar{N}}$ 
has only $D\times D$ independent components. 
In fact, by using a certain matrix $s_{mn}$, $\cS_{MN}$ can be expressed as 
\begin{align}
 (\cS_{MN})= \begin{pmatrix}
 2\,G_{(m|k}\,s^{[kl]} \,B_{l|n)} - s_{(mn)} 
 - B_{mk}\,s^{(kl)}\,B_{ln}\quad & B_{mk}\,s^{(kn)} - G_{mk}\,s^{[kn]} \\
 s^{[mk]}\,G_{kn} -s^{(mk)}\,B_{km}\quad & s^{(mn)}
 \end{pmatrix} \,.
\label{eq:cS-matrix}
\end{align}
Then, $\cS_{MN}=0$ is equivalent to $s_{mn}=0$\,. 
When we choose the canonical section $\partial_M=(\partial_m,\,0)$ 
and adopt the parameterizations in \eqref{eq:cH-parametrization}, 
we can express $s_{mn}$ as
\begin{align}
 s_{mn} = R_{mn}-\frac{1}{4}\,H_{mpq}\,H_n{}^{pq} + 2 D_m \partial_n \Phi 
 - \frac{1}{2}\,D^k H_{kmn} + \partial_k\Phi\,H^k{}_{mn} \,. 
\end{align}
If we decompose $s_{mn}$ into the symmetric part $s_{(mn)}$ 
and the anti-symmetric part $s_{[mn]}$, the well-known beta functions \cite{Callan:1985ia} in the conventional string sigma model 
are reproduced \footnote{Note that the beta function 
in Tseylin's double sigma model \cite{Tseytlin:1990nb,Tseytlin:1990va}
\begin{align*}
 S = \frac{1}{4\pi\alpha'}\int_\Sigma\rmd^2\sigma\,\bigl(\eta_{MN}\,\partial_1 X^M\,
 \partial_0 X^N - \cH_{MN}\,\partial_1 X^M\,\partial_1 X^N \bigr) \,, 
\end{align*}
is also computed in \cite{Berman:2007xn,Berman:2007yf,Copland:2011yh,Copland:2011wx} 
to have the form, $\beta_{MN}=\cS_{MN}$, 
unifying $\beta^G_{mn}$ and $\beta^B_{mn}$ in a covariant manner. }
\begin{align}
 s_{(mn)}&=\beta^G_{mn} \equiv R_{mn}-\frac{1}{4}\,H_{mpq}\,H_n{}^{pq} + 2 D_m \partial_n \Phi \,,
\label{eq:beta-G}
\\
 s_{[mn]}&=\beta^B_{mn} \equiv - \frac{1}{2}\,D^k H_{kmn} + \partial_k\Phi\,H^k{}_{mn} \,. 
\label{eq:beta-B}
\end{align}
Thus the flatness condition of the generalized Ricci tensor $\cS_{MN}=0$ is equivalent to
\begin{align}
 \beta^G_{mn}=0 \,,\qquad \beta^B_{mn}=0 \,.
\end{align}
On the other hand, $\cS$ has the form
\begin{align}
 \cS = R + 4\,D^m \partial_m \Phi - 4\,\abs{\partial \Phi}^2 - \frac{1}{2}\,\abs{H_3}^2 \,,
\label{eq:beta-Phi}
\end{align}
and $\cS=0$ is equivalent to the vanishing beta function for the dilaton $\beta^\Phi=0$\,. 
To summarize, it has been shown that $\cG_{MN} = 0$ is equivalent to 
\begin{align}
 \beta^G_{mn}=\beta^B_{mn}=\beta^\Phi=0\,. 
\end{align}
These equations are nothing but the equations of motion for the NS-NS sector 
of the conventional supergravity. 

\medskip 

As a side remark, it will be interesting to note that $\beta^\Phi$ 
becomes a constant upon using equations of motion $\cS_{MN}=0$\,, 
as emphasized in \cite{Callan:1985ia}. 
Indeed, by using the differential Bianchi identity and $\cS_{MN}=0$\,, 
it is easy to show that 
\begin{align}
 \partial_M \cS = 2\,\cH_{MN}\,\nabla_K \cS^{KN} = 0\,. 
\end{align}

\medskip 

In the next section, we will consider a modification of the standard DFT introduced here 
by relaxing the condition (4) for the $T$-duality invariant dilaton $d(x)$\,.

\section{The generalized equations from modified DFT}
\label{sec:DFT-generalized}

In this section, we will consider a modification of the generalized connection 
by relaxing the condition (4) associated with the DFT dilaton $d(x)$\,. 
This leads to the modification of the generalized Ricci tensors. 
We here study a theory, referred to as a modified double field theory (mDFT), 
whose equations of motion are given by the modified generalized Ricci flatness conditions. 

\medskip 

The goal of this section is to derive the generalized equations in the NS-NS sector, 
\begin{align}
\begin{split}
 &R_{mn}-\frac{1}{4}\,H_{mpq}\,H_n{}^{pq} + D_m X_n + D_n X_m = 0\,,
\\
 &\frac{1}{2}\,D^k H_{kmn} - \bigl(X^k H_{kmn} + D_m X_n - D_n X_m \bigr)=0\,,
\\
 &R-\frac{1}{2}\,\abs{H_3}^2 + 4\,D_m X^m - 4\,X^m X_m=0\,,
\end{split}
\label{eq:EOM-gen-sugra}
\end{align}
from the equations of motion in mDFT. 
Namely, we show that the generalized equations are embedded into the mDFT. 

\medskip

We also discuss several aspects in mDFT; the global $\OO(D,D)$ symmetry, modification of the string charge, and the action. 

\subsection{Modified generalized Ricci tensor}
\label{sec:modified-Ricci}

Let us recall that the generalized covariant derivative has been defined by requiring four conditions in section \ref{sec:generalized-conection}. 

\medskip 

Our central idea here is to make a modification of the condition (4) 
which determines the contracted connection $\Gamma_M=\Gamma_K{}^K{}_M$\,. 
The condition (4) was originally required for the covariant divergence of 
an arbitrary generalized vector $V^M$ multiplied by the measure $\Exp{-2d}$ 
to be a total derivative,
\begin{align}
 \Exp{-2d}\nabla_M V^M = \partial_M \bigl(\Exp{-2d} V^M\bigr)\,.
\label{eq:total-div}
\end{align} 
In the following, we will consider the following modification of this condition:
\begin{align}
 \bnabla_M d = \partial_M d + \frac{1}{2}\,\bGamma_K{}^K{}_M = -\bX_M \,, 
\end{align}
by introducing an extra generalized vector field $\bX_M$\,. 
Here and hereafter, we denote the $\bX$-modified generalized covariant derivative 
and the generalized connection by $\bnabla$ and $\bGamma$\,. 
For the modified generalized connection, the relation in \eqref{eq:total-div} is also deformed;
\begin{align}
 \Exp{-2d} \bnabla_M V^M = \partial_M \bigl(\Exp{-2d} V^M\bigr) 
 - 2\,\Exp{-2d}\bX_M\,V^M \,. 
\label{eq:div}
\end{align}
As we will discuss in section \ref{sec:action}, this deformation makes it difficult 
to derive the generalized Ricci flatness conditions as the equations of motion. 

\medskip 

By using the expression \eqref{eq:connection-components} and
\begin{align}
 \bGamma_M = -2\,\bigl(\partial_M d + \bX_M\bigr) \,, 
\end{align}
it is easy to calculate the difference in the generalized connection,
\begin{align}
 \delta_\bX \Gamma_{MNK}\equiv \bGamma_{MNK}-\Gamma_{MNK}\,. 
\end{align}
Indeed, we obtain
\begin{align}
 \delta_\bX \Gamma_{MNK} 
  = - \frac{4}{D-1}\,\big(P_{M[N}\,P_{K]}{}^L +\bar{P}_{M[N}\,\bar{P}_{K]}{}^L\big)\,\bX_L \,.
\end{align}
In order to calculate the modifications of the generalized Ricci tensors, 
let us employ the formula for the variation of 
the generalized Riemann tensor \cite{Hohm:2011si},%
\footnote{The generalized connection 
$\Gamma_{MNK}$ here corresponds to $\Gamma_{MKN}=-\Gamma_{MNK}$ 
of \cite{Hohm:2011si}.}
\begin{align}
 \delta_\bX S_{MNKL} &= \nabla_{[M} \delta_\bX \Gamma_{N]KL} 
 + \nabla_{[K}\delta_\bX \Gamma_{L]MN} 
\nn\\
 &\quad 
 - \delta_\bX\Gamma_{[M|PL}\,\delta_\bX\Gamma_{|N]K}{}^P
 - \delta_\bX \Gamma_{[K|PN}\,\delta_\bX\Gamma_{|L]M}{}^P 
 - \frac{1}{2}\,\delta_\bX\Gamma_{PMN}\,\delta_\bX\Gamma^P{}_{KL} \,,
\end{align}
under a shift in the generalized connection
\begin{align}
 \Gamma_{MNK}\to \bGamma_{MNK}=\Gamma_{MNK}+\delta_\bX \Gamma_{MNK}\,.
\end{align}
This formula leads to the variation of the generalized Ricci tensor: 
\begin{align}
 \delta_\bX \cS_{MN} &= -2\,\bigl(P_M{}^K\,\bar{P}_N{}^L+\bar{P}_M{}^K\,P_N{}^L \bigr)\,
 \Bigl[-\nabla_{(M} \delta_\bX \Gamma_{N)} - \nabla^K \delta_\bX \Gamma_{(MN)K} 
\nn\\
 &\qquad\qquad
 - \delta_\bX \Gamma_{(M|}{}^{KL}\,\delta_\bX\Gamma_{L|N)K} 
 - \delta_\bX \Gamma_K \,\delta_\bX\Gamma_{(MN)}{}^K 
 - \frac{1}{2}\,\delta_\bX\Gamma_{KM}{}^L \,\delta_\bX\Gamma^K{}_{NL} \Bigr]
\nn\\
 &= -4\,\bigl(P_{MK}\,\bar{P}_{NL} + \bar{P}_{MK}\,P_{NL})\,\nabla^{(K} \bX^{L)} \,.
\end{align}
Note here that only the first term has non-vanishing contribution 
because $\delta_\bX\Gamma_{MNK}$ has only $\bar{M}\bar{N}\bar{K}$ 
and $\ubar{M}\ubar{N}\ubar{K}$ components and the projections remove all the other terms. 
A similar calculation also leads to 
\begin{align}
 \delta_\bX \cS = 4\,\cH_{MN}\,\nabla^M\bX^N - 4\,\cH_{MN}\,\bX^M\,\bX^N \,. 
\label{eq:delta-X-cS}
\end{align}
Then, by introducing a dual generalized vector 
\begin{align}
 \bY^M\equiv \cH^{MN}\,\bX_N\,, 
\end{align}
the following expressions are obtained 
\begin{align}
 \delta_\bX \cS_{MN} &= 4\,\bigl(P_{MK}\,\bar{P}_{NL} 
 - \bar{P}_{MK}\,P_{NL})\,\nabla^{[K} \bY^{L]} = -\gLie_\bY \cH_{MN} \,, 
\\
 \delta_\bX \cS &= 4\, \nabla_M \bY^M - 4\,\cH_{MN}\,\bY^M\,\bY^N \,. 
\end{align}

\medskip 

In summary, by using the expressions of $\cS_{MN}$ and $\cS$, the modified generalized Ricci tensors, $\bS_{MN}$ and $\bS$, are written as 
\begin{align}
 \bS_{MN}&= \cS_{MN} - \gLie_\bY \cH_{MN}\,,
\\
 \bS&= \frac{1}{8}\,\cH^{MN}\,\partial_M \cH^{KL}\partial_N \cH_{KL} 
 -\frac{1}{2}\,\cH^{KL}\,\partial_L\cH^{MN}\,\partial_N \cH_{KM} 
 + 4\partial_M d\,\partial_N\cH^{MN} 
\nn\\
 &\quad -4\,\cH^{MN}\,\partial_M d\,\partial_N d
 - \partial_M\partial_N\cH^{MN} 
 +4\,\cH^{MN}\,\partial_M\partial_N d 
\nn\\
 &\quad +4\,\nabla_M \bY^M -4\,\cH_{MN}\,\bY^M \,\bY^N \,.
\end{align}
It is easy to see that $\bS$ is derived from $\cS$ by making a simple replacement,
\begin{align}
 \partial_M d\to \partial_M d + \bX_M\,.
\end{align}
This must be also true for $\bS_{MN}$ since the dilaton dependence in $\bS_{MN}$ appears only with the combination, $\partial_M d + \bX_M$. 

\subsection{Equations of motion as the generalized Ricci flatness conditions}
\label{sec:EOM-mDFT}

The aim of this subsection is to show that the modified generalized Ricci flatness conditions,
\begin{align}
 \bS_{MN} = 0 \,,\qquad \bS = 0 \,,
\label{eq:generalized-Ricci-flat}
\end{align}
or equivalently,
\begin{align}
 \cS_{MN} = \gLie_\bY \cH_{MN} \,,\qquad 
 \cS = - 4\,\nabla_M \bY^M + 4\,\cH_{MN}\,\bY^M \bY^N \,,
\end{align}
lead to the equations of motion in the generalized supergravity \eqref{eq:EOM-gen-sugra}. 

\medskip

In order to reproduce \eqref{eq:EOM-gen-sugra}, we choose the canonical section 
$(\partial_M)=(\partial_m,\,0)$, and adopt the parameterizations \eqref{eq:cH-parametrization}. 
We also parameterize the null generalized Killing vector as
\begin{align}
 \bX^M \equiv \begin{pmatrix} \delta^m_n & 0 \\ B_{mn} & \delta_m^n \end{pmatrix} 
 \begin{pmatrix} I^n \\ U_n \end{pmatrix} = \begin{pmatrix} I^m \\ U_m + B_{mn}\, I^n
 \end{pmatrix} \,. 
\label{eq:X-parameterization}
\end{align}
With this parameterization, $I^m$ and $U_m$ are a contravariant and a covariant vector, 
respectively, and invariant under the $B$-field gauge transformations. 
The explicit form of the dual generalized vector becomes
\begin{align}
 \bY^M = \cH^M{}_N\,\bX^N = \begin{pmatrix} \delta^m_n & 0 \\ B_{mn} & 
 \delta_m^n \end{pmatrix} \begin{pmatrix} U^n \\ I_n \end{pmatrix}
 = \begin{pmatrix} U^m \\ I_m + B_{mn}\,U^n \end{pmatrix} \,.
\end{align}
Then, the generalized Lie derivative $\gLie_\bY \cH_{MN}$, 
which has the components $(\gLie_\bY \cH)_{MN}=(\gLie_\bY \cH)_{\ubar{M}\bar{N}}
+(\gLie_\bY \cH)_{\bar{M}\ubar{N}}$\,, becomes a matrix of 
the form \eqref{eq:cS-matrix} with $s_{mn}$ replaced by
\begin{align}
 \ell_{mn} = - \bigl(D_m U_n +D_n U_m + U^k\,H_{kmn} + D_m I_n - D_n I_m \bigr) \,.
\end{align}
From \eqref{eq:beta-G}, \eqref{eq:beta-B}, and \eqref{eq:beta-Phi}, 
the modified generalized Ricci flatness conditions \eqref{eq:generalized-Ricci-flat} 
can be expressed as
\begin{align}
 \bs_{(mn)}=0\,,\qquad \bs_{[mn]} = 0 \,,\qquad \bS=0 
\end{align}
with
\begin{align}
\begin{split}
 \bs_{(mn)}&\equiv s_{(mn)}+\ell_{(mn)} = R_{mn}-\frac{1}{4}\,H_{mpq}\,H_n{}^{pq} 
 + 2 D_m \partial_n \Phi + D_m U_n +D_n U_m \,,
\\
  \bs_{[mn]}&\equiv s_{[mn]}+\ell_{[mn]} = -\frac{1}{2}\,D^k H_{kmn} + \partial_k\Phi\,H^k{}_{mn}
 + U^k\,H_{kmn} + D_m I_n - D_n I_m \,,
\\
 \bS=&R + 4\,D^m \partial_m \Phi - 4\,\abs{\partial \Phi}^2 - \frac{1}{2}\,\abs{H_3}^2 
 - 4\,\bigl(I^m I_m+U^m U_m + 2\,U^m\,\partial_m \Phi - D_m U^m\bigr) \,. 
\end{split}
\end{align}

\medskip 

Although these equations are not equivalent to \eqref{eq:EOM-gen-sugra} yet, 
we can show the equivalence by further assuming that the generalized vector,
\begin{align}
 \bX^M=-\bnabla^Md \,,
\end{align}
is a null generalized Killing vector;
\begin{align}
 \eta_{MN}\,\bX^M \bX^N=0\,,\qquad 
 \gLie_\bX \cH_{MN}=0\,,\qquad \gLie_\bX d=0\,. 
\end{align}
The null property gives rise to 
\begin{align}
 I^m\,U_m=0 \,,
\end{align}
and the generalized Killing property leads to the following relations: 
\begin{align}
 D_m I_n + D_n I_m =0 \,,\qquad 
 I^k\,H_{kmn} + D_m U_n - D_n U_m \,,\qquad 
 I^m\,\partial_m \Phi = 0\,. 
\end{align}
These requirements lead to the equations of motion 
\eqref{eq:EOM-gen-sugra} with identifications 
\begin{align}
 X_m = I_m + Z_m\,, \qquad Z_m = \partial_m \Phi + U_m\,.
\end{align}
The null property and the generalized Killing property can be summarized as
\begin{align}
 D_m I_n + D_n I_m =0\,, \qquad 
 I^k\,H_{kmn} + D_m Z_n - D_n Z_m =0\,,\qquad I^m\,Z_m = 0\,. 
\label{eq:conditions}
\end{align}
Note that, as we discuss in section \ref{sec:string-charge}, the \emph{null} and the \emph{generalized Killing} properties are important in order to define a ``fundamental string charge'' in mDFT. 
This result indicates that the kappa-symmetry constraints play an important role in defining the ``string charge.'' 

\medskip

Note also that the equations of motion can go back to the original ones if and only if
\begin{align}
 \gLie_\bY \cH_{MN} = 0 \,,\qquad \cH_{MN}\,\bY^M \bY^N - \nabla_M \bY^M = 0 \,,
\end{align}
are satisfied. 

\subsection{Global $\OO(D,D)$ transformations in mDFT}
\label{sec:DFT-mDFT}

In this subsection, we consider global transformations which map a solution to other solutions. 

\medskip

First of all, let us consider global $\OO(D,D)$ transformations in DFT. 
An arbitrary global $\OO(D,D)$ transformation can be generated by the following three transformations:
\begin{align}
 \text{\underline{$\GL(D)$:}}\qquad & 
 \Lambda(a)_M{}^N\equiv 
 \begin{pmatrix}
  a_m{}^n & 0 \\ 0 & (a^{-\rmT})^m{}_n
 \end{pmatrix}\qquad 
 \bigl[\,a_m{}^n\in \GL(D)\,\bigr]\,,
\\
 \text{\underline{$B$-shift:}}\qquad & 
 \Lambda(b)_M{}^N\equiv 
 \begin{pmatrix}
  \delta_m^n & b_{mn} \\ 0 & \delta^m_n
 \end{pmatrix}\qquad \bigl[\,b_{mn}=-b_{nm}\,\bigr]\,,
\\
 \text{\underline{$T$-duality:}}\qquad & 
 \Lambda(k)_M{}^N\equiv 
 \begin{pmatrix}
  \delta_m^n - \delta_m^k\,\delta^n_k & \delta_m^k\,\delta_n^k \\ \delta^m_k\,\delta^n_k & \delta^m_n - \delta^m_k\,\delta_n^k
 \end{pmatrix} \,. 
\end{align}
Here, ``$T$-duality'' represents the conventional $T$-duality transformation along $x^k$ direction and the index $k$ in the last expression is not summed. 
Under a general global $\OO(D,D)$ transformation $\Lambda_M{}^N$\,, 
the generalized coordinates, the derivatives, and the generalized metric are defined so as to transform covariantly,
\begin{align}
 x^M\to x'^M= \Lambda^M{}_N\, x^N\,, \quad 
 \partial_M\to \partial'_M= \Lambda_M{}^N\,\partial_N \,,\quad 
 \cH_{MN}\to \cH'_{MN}=\Lambda_M{}^K\,\Lambda_N{}^L\,\cH_{KL}\,. 
\label{eq:coord-rotated}
\end{align}
Here the indices are raised or lowered with $\eta_{MN}$ as usual. 
In addition, the transformation rule for the DFT dilaton $d(x)$ is given as follows. 
Since $\Exp{-2d}$ is a scalar density, it is transformed under a global $\GL(D)$ transformations as
\begin{align}
 \Exp{-2d} ~~\to~~ \Exp{-2d'} = \abs{\det(a)}\,\Exp{-2d} \,.
\end{align}
On the other hand, it is invariant under the $B$-shifts and $T$-duality transformations. 
Then, since $\abs{\det(a)}$ is constant, $\partial_M d$ is transformed as a generalized vector,
\begin{align}
  \partial_M d ~~\to~~ \partial'_M d' = \Lambda_M{}^N\,\partial_N d \,, 
\end{align}
under global $\OO(D,D)$ transformations. 
The generalized Ricci tensors, $\cS_{MN}$ and $\cS$, are polynomials in
\begin{align}
 \{\cH_{MN}\,,\ \partial_K \cH_{MN}\,,\ \partial_K\partial_L \cH_{MN}\,,\ 
 \partial_M d \,,\ \partial_M \partial_N d\} 
\end{align}
with the indices contracted with $\eta_{MN}$, and hence it is useful to write $\cS_{MN}$ and $\cS$ in terms of polynomials $\cP_{MN}$ and $\cP$ as
\begin{align}
\begin{split}
 \cS_{MN} &\equiv \cP_{MN}\bigl(\cH_{MN},\, \partial_K \cH_{MN},\, \partial_K\partial_L \cH_{MN},\, \partial_M d,\, \partial_M \partial_N d\bigr)\,,
\\
 \cS &\equiv \cP\bigl(\cH_{MN},\, \partial_K \cH_{MN},\, \partial_K\partial_L \cH_{MN},\, \partial_M d,\, \partial_M \partial_N d\bigr)\,. 
\end{split}
\end{align}
We then have the following transformation rule for $\cS_{MN}$ and $\cS$\,:
\begin{align}
 \cS'_{MN} = \Lambda_M{}^K\,\Lambda_N{}^L\,\cS_{KL} \,, \qquad
 \cS' = \cS \,, 
\label{eq:cS-odd}
\end{align}
where
\begin{align}
\begin{split}
 \cS'_{MN} &\equiv \cP_{MN}\bigl(\cH'_{MN},\, \partial'_K \cH'_{MN},\, \partial'_K\partial'_L \cH'_{MN},\, \partial'_M d',\, \partial'_M \partial'_N d'\bigr)\,,
\\
 \cS' &\equiv \cP\bigl(\cH'_{MN},\, \partial'_K \cH'_{MN},\, \partial'_K\partial'_L \cH'_{MN},\, \partial'_M d',\, \partial'_M \partial'_N d'\bigr)\,.
\end{split}
\end{align}
This shows that under the global $\OO(D,D)$ transformation, the DFT action is invariant and the equations of motion are covariant. 
In the case of \eqref{eq:coord-rotated}, however, everything is rotated and nothing is changed physically. 

\medskip 

Next, let us recall that the strong constraint can be satisfied by choosing the canonical section $(\partial_M)=(\partial_m,\,0)$, up to global $\OO(D,D)$ rotations of generalized coordinates. 
Once the canonical section has been fixed, we will not change doubled coordinates any longer but perform a different kind of global $\OO(D,D)$ transformation which acts only on fields, $\cH_{MN}$ and $d(x)$, on the canonical section:
\begin{align}
 x^M~ \to ~ x'^M\,,\qquad 
 \cH_{MN} ~\to ~\cH'_{MN}=\Lambda_M{}^K\,\Lambda_N{}^L\,\cH_{KL}\,, \qquad 
 \partial_M d ~\to ~ \partial_M d' = \partial_M d \,. 
\end{align}
After performing this transformation, the equations of motion become
\begin{align}
 \tilde{\cS}_{MN} =0 \,,\qquad \tilde{\cS} = 0\,,
\end{align}
where
\begin{align}
\begin{split}
 \tilde{\cS}_{MN} &\equiv \cP_{MN}\bigl(\cH'_{MN},\, \partial_K \cH'_{MN},\, \partial_K\partial_L \cH'_{MN},\, \partial_M d',\, \partial_M \partial_N d'\bigr) \,,
\\
 \tilde{\cS} &\equiv \cP\bigl(\cH'_{MN},\, \partial_K \cH'_{MN},\, \partial_K\partial_L \cH'_{MN},\, \partial_M d',\, \partial_M \partial_N d'\bigr) \,. 
\end{split}
\end{align}
Then, even if we start from a solution of DFT, $\bigl\{\cH_{MN}(x),\,d(x)\bigr\}$, the transformed configuration, $\bigl\{\cH'_{MN}(x),\,d'(x)\bigr\}$, may not satisfy the equations of motion, because the original equations of motion, $\cS_{MN}=\cS=0$, and the equations of motion for the transformed solution, $\tilde{\cS}_{MN} = \tilde{\cS} = 0$, are not equivalent (although $\cS_{MN}=\cS=0$ and $\cS'_{MN}=\cS'=0$ are equivalent from \eqref{eq:cS-odd}). 
However, if we consider a special case, in which there exist isometries, 
$\cH_{MN}(x)$ and $d(x)$ are independent of the corresponding coordinates, denoted by $(y^i)$ $(i=1,\dotsc,N)$\,. 
Then, by decomposing the physical coordinates as $(x^m)=(x^\mu,\,y^i)$ ($\mu=1,\dotsc,D-N$), 
the derivative takes the following form:
\begin{align}
 \bigl(\partial_M)=(\partial_\mu,\,\partial_i,\,\tilde{\partial}^\mu,\,
 \tilde{\partial}^i\bigr)^\rmT =(\partial_\mu,\,0,\,0,\,0\bigr)^\rmT\,. 
\end{align}
In the case with isometries, under global $\OO(N,N)$ transformations of the form,
\begin{align}
 \bigl(\Lambda_M{}^N\bigr) 
 = \begin{pmatrix}
 \delta_\mu{}^\nu & 0 & 0 & 0 \\
 0 & \Lambda_i{}^j & 0 & \Lambda_{ij} \\
 0 & 0 & \delta^\mu{}_\nu & 0 \\
 0 & \Lambda^{ij} & 0 & \Lambda^i{}_j 
 \end{pmatrix} \,,
\label{eq:ONN}
\end{align}
the derivative $\partial_M$ is the same as the would-be transformed derivative,
\begin{align}
 \partial'_M \equiv \Lambda_M{}^N\,\partial_N =(\partial_\mu,\,0,\,0,\,0\bigr)^\rmT \,.
\end{align}
We then obtain 
\begin{align}
\begin{split}
 \tilde{\cS}_{MN} &= \cP_{MN}\bigl(\cH'_{MN},\, \partial'_K \cH'_{MN},\, \partial'_K\partial'_L \cH'_{MN},\, \partial'_M d',\, \partial'_M \partial'_N d'\bigr) 
 = \cS'_{MN} \,,
\\
 \tilde{\cS} &= \cP\bigl(\cH'_{MN},\, \partial'_K \cH'_{MN},\, \partial'_K\partial'_L \cH'_{MN},\, \partial'_M d',\, \partial'_M \partial'_N d'\bigr) = \cS' \,.
\end{split}
\end{align}
Recalling $\cS_{MN}=\cS=0$ is equivalent to $\cS'_{MN}=\cS'=0$, we obtain $\tilde{\cS}_{MN}=\tilde{\cS}=0$ and $\bigl\{\cH'_{MN}(x),\,d'(x)\bigr\}$ satisfies the equations of motion of DFT. 
Namely, in the presence of $N$ isometries, $\OO(N,N)$ transformations \eqref{eq:ONN} map a solution to other solutions of DFT. 
These are nothing but the $\OO(N,N)$ $T$-duality transformations known in the conventional supergravity. 

\medskip 

Let us next consider a more general class of DFT solutions in which $\cH_{MN}$ is independent 
of $y^i$ but the DFT dilaton has a linear dependence on $y^i$, namely,
\begin{align}
 \cH_{MN}(x,y)=\cH_{MN}(x)\,,\qquad 
 d(x,y)=d_0(x)+c_i\,y^i \qquad (c_i: ~\mbox{constants})\,.
\label{eq:linear-solution}
\end{align}
We show that the above $y^i$-dependent solution of DFT is mapped 
to solutions of mDFT under global $\OO(N,N)$ transformations. 

\medskip

Starting from the solution \eqref{eq:linear-solution} of DFT, $\{\cH_{MN},\,d\}$, let us consider an $\OO(N,N)$ transformation \eqref{eq:ONN} and denote the transformed configuration by $\{\cH'_{MN},\,d'\}$\,. 
As explained above, due to the presence of the isometries, 
$\partial_K \cH'_{MN}$ and $\partial_K\partial_L \cH'_{MN}$ can be rewritten 
as $\partial'_K \cH'_{MN}$ and $\partial'_K\partial'_L \cH'_{MN}$\,, as in the previous case. 
On the other hand, the non-isometry of $d$ leads to
\begin{align}
 \partial_M d'\neq \partial'_M d' \quad \mbox{for}~~ c_i\neq 0\,,
\end{align}
as we can see from the explicit expression,
\begin{align}
 \partial_M d' 
 = \begin{pmatrix} \partial_\mu d' \\ \partial_i d' \\ \tilde{\partial}^\mu d' \\ \tilde{\partial}^i d' \end{pmatrix} 
 = \begin{pmatrix} \partial_\mu d \\ c_i \\ 0\\ 0 \end{pmatrix}\,,\qquad
 \partial'_M d' 
 = \Lambda_M{}^N\,\partial_N d'
 = \begin{pmatrix} \partial_\mu d \\ \Lambda_i{}^j\,c_j \\ 0 \\ \Lambda^{ij}\,c_j \end{pmatrix}\,. 
\end{align}
Again, $\cS_{MN}=\cS=0$ is equivalent to $\cS'_{MN}=\cS'=0$ but not equivalent to $\tilde{\cS}_{MN}=\tilde{\cS}=0$, and the transformed configuration $\{\cH'_{MN},\,d'\}$ is not a solution of DFT (in general). 

\medskip 

We now expand our framework from DFT to mDFT. 
Suppose that the initial configuration given by \eqref{eq:linear-solution} and $\bX^M=0$ is a solution of DFT;
\begin{align}
 \bS_{MN} = \cS_{MN} = 0 \,,\qquad \bS = \cS = 0 \,. 
\end{align}
Since $\bS_{MN}$ and $\bS$ can be obtained from $\cS_{MN}$ and $\cS$ with the replacement,
\begin{align}
 \partial_M d ~~\to \partial_M d + \bX_M \,,
\end{align}
we have
\begin{align}
\begin{split}
 \bS_{MN} &= \cP_{MN}\bigl(\cH_{MN},\, \partial_K \cH_{MN},\, \partial_K\partial_L \cH_{MN},\, \partial_M d+\bX_M,\, \partial_M (\partial_N d+\bX_N)\bigr) \,,
\\
 \bS &= \cP\bigl(\cH_{MN},\, \partial_K \cH_{MN},\, \partial_K\partial_L \cH_{MN},\, \partial_M d+\bX_M,\, \partial_M (\partial_N d+\bX_N)\bigr) \,. 
\end{split}
\end{align}
Again, $\bS_{MN}=\bS=0$ is equivalent to $\bS'_{MN}=\bS'=0$ if $\bS'_{MN}$ and $\bS'$ take the forms,
\begin{align}
\begin{split}
 \bS'_{MN} &\equiv \cP_{MN}\bigl(\cH'_{MN},\, \partial'_K \cH'_{MN},\, \partial'_K\partial'_L \cH'_{MN},\, \partial_M d'+\bX'_M,\, \partial'_M (\partial_N d'+\bX'_N)\bigr)\,,
\\
 \bS' &\equiv \cP\bigl(\cH'_{MN},\, \partial'_K \cH'_{MN},\, \partial'_K\partial'_L \cH'_{MN},\, \partial_M d'+\bX'_M,\, \partial'_M (\partial_N d'+\bX'_N)\bigr)\,,
\end{split}
\end{align}
where the combination $\partial_M d'+\bX'_M$ is defined by
\begin{align}
 \partial_M d'+\bX'_M \equiv \Lambda_M{}^N \, \bigl(\partial_N d +\bX_N \bigr) \,.
\label{eq:dd+X}
\end{align}
Then, if one could find $d'$ and $\bX'_M$ which satisfy \eqref{eq:dd+X}, and both $\cH'_{MN}$ and $\partial_M d'+\bX'_M$ are independent of $y^i$, the configuration $\{\cH'_{MN},\,d',\,\bX'_M\}$ is a solution of mDFT, since we have
\begin{align}
\begin{split}
 \tilde{\bS}_{MN} &\equiv \cP_{MN}\bigl(\cH'_{MN},\, \partial_K \cH'_{MN},\, \partial_K\partial_L \cH'_{MN},\, \partial_M d'+\bX'_M,\, \partial_M (\partial_N d'+\bX'_N)\bigr)=\bS'_{MN}\,,
\\
 \tilde{\bS} &\equiv \cP\bigl(\cH'_{MN},\, \partial_K \cH'_{MN},\, \partial_K\partial_L \cH'_{MN},\, \partial_M d'+\bX'_M,\, \partial_M (\partial_N d'+\bX'_N)\bigr)=\bS'\,,
\end{split}
\end{align}
and $\bS_{MN}=\bS=0$ becomes equivalent to $\tilde{\bS}_{MN}=\tilde{\bS}=0$\,. 
In a special case where $\bX'_M=0$, the transformed configuration satisfies also the equations of motion of DFT. 

\medskip

Now, let us find out the transformed solution $d'$ and $\bX'_M$ from the initial solution \eqref{eq:linear-solution} and $\bX_M=0$\,. 
In this case, the right-hand side of \eqref{eq:dd+X} becomes
\begin{align}
 \Lambda_M{}^N \, \bigl(\partial_N d +\bX_N \bigr) 
 = \Lambda_M{}^N \,\partial_N d 
 = (\partial_\mu d,\, \Lambda_i{}^j\,c_j,\, 0,\, \Lambda^{ij}\,c_j)^\rmT\,,
\end{align}
and we need to find a pair $\{d',\,\bX'_M\}$ which satisfies
\begin{align}
 (\partial_M d' + \bX_M')(x) = \bigl(\partial_\mu d \,,\ \Lambda_i{}^j\,c_j \,,\ 0 \,, \ \Lambda^{ij}\,c_j\bigr)^\rmT \,.
\label{eq:DFT-GS}
\end{align}
Note here that this quantity is independent of $y^i$\,. 
Although there is an ambiguity in decomposing $\partial_M d'+\bX'_M$ into $\partial_M d'$ and $\bX'_M$\,,%
\footnote{This ambiguity is not important at the level of the equations of motion 
because $d'$ appears only through the combination $\partial_M d'+\bX'_M$\,. 
However, if we consider for example the action, $d'$ enters directly 
and we need to treat the ambiguity carefully.}
a convenient choice is
\begin{align}
 d'(x,y) = d(x,y) + (\Lambda_i{}^j-\delta_i^j)\,c_j\, y^i \,,\qquad 
 \bX'_M= \bigl(0 \,,\ 0 \,,\ 0 \,, \ \Lambda^{ij}\,c_j\bigr)^\rmT \,.
\end{align}
These, together with $\cH'_{MN}=\Lambda_M{}^K\,\Lambda_N{}^L\,\cH_{KL}$ satisfies the equations of motion of mDFT. 

\medskip 

In order to rewrite the above solution in terms of the generalized supergravity, 
let us parameterize $\bX^M$ as \eqref{eq:X-parameterization}. 
Then, we obtain
\begin{align}
 I'^\mu = 0\,,\qquad I'^i=\Lambda^{ij}\,c_j\,,\qquad 
 U'_m= - B'_{mn}\,I'^n \,. 
\end{align}
If the conventional dilaton $\Phi'(x,y)$ is defined as usual, 
$\Exp{-2d'}=\Exp{-2\Phi'}\sqrt{\abs{G'}}$\,, 
and also define vector fields $Z'_m$ and $X'_m$ by
\begin{align}
 Z'_m(x) \equiv \partial_m \Phi'-B'_{mn}\,I'^n\,,\qquad 
 X'_m(x) \equiv I'_m + Z'_m \,,
\end{align}
the configuration $\{G'_{mn}(x),\,B'_{mn}(x),\,X'_m(x)\}$ 
satisfies the generalized equations \eqref{eq:EOM-gen-sugra}. 
Note also that $I'^m$ and $Z'_m$ defined above satisfies conditions 
\eqref{eq:conditions} for arbitrary $\Lambda_M{}^N$ and $c^i$. 
The only non-trivial relation, 
\begin{align}
 I^m\,Z_m = I^j\,\partial_j \Phi'=0
\end{align}
can be shown by using an $\OO(N,N)$ property, $\Lambda_k{}^{(i|}\,\Lambda^{k|j)}=0$\,. 

\medskip 

In summary, we have shown that for a solution of DFT given by
\begin{align}
 \bigl\{\cH_{MN}(x),\ d(x,y)=d_0(x)+c_i\,y^i\ \bigl(= \Phi(x,y) - \ln \abs{\det G_{mn}}^{1/4}\bigr)\bigr\}\,,
\label{eq:sol-DFT}
\end{align}
there exists a solution of mDFT,
\begin{align}
 \bigl\{\cH'_{MN}(x),\ d'(x,y)=d_0(x) + \Lambda_i{}^j\,c_j\, y^i\,,\ \bX'_M = \bigl(0 \,,\ 0 \,,\ 0 \,, \ \Lambda^{ij}\,c_j\bigr)^\rmT \bigr\}\,, 
\label{eq:sol-mDFT}
\end{align}
or equivalently a solution of the generalized equations
\begin{align}
 \{G'_{mn}(x),\,B'_{mn}(x),\,X'_m(x)\} \,, 
\end{align}
where the transformation rule is given as follows:
\begin{align}
\begin{split}
 \cH'_{MN}(x)&=\Lambda_M{}^K\,\Lambda_N{}^L\,\cH_{KL}(x) \qquad
 \bigl[\,\Lambda_M{}^N\in \OO(N,N)\,\bigr] \,,
\\
 X'_m(x) &= I'_m + Z'_m = \partial_m \Phi' + (G'-B')_{mn}\,I'^n \,, 
\\
 \Phi'(x,y)&=\Phi(x,y) + \ln\Bigl\lvert \frac{\det G'_{mn}}{\det G_{mn}}\Bigr\rvert^{1/4} 
+ (\Lambda_i{}^j-\delta_i^j)\,c_j\, y^i \,, \qquad 
 I'^m = (0,\, \Lambda^{ij}\,c_j)\,.
\end{split}
\label{eq:formula}
\end{align}

\subsubsection*{Example}

As a simple example, let us consider a solution of the conventional supergravity \cite{AFHRT},
\begin{align}
 \rmd s^2 = \Exp{2a(x)}\bigl(\rmd y+A_\mu(x)\,\rmd x^\mu\bigr)^2 
 + g_{\mu\nu}(x)\,\rmd x^\mu\,\rmd x^\nu\,,\quad 
 B_2=0\,,\quad 
 \Phi = c\,y + f(x) \,. 
\end{align}
Let us perform a $T$-duality along $y$ direction, i.e., $\Lambda_y{}^y = 0$ and $\Lambda^{yy}=1$\,. 
Then the metric and $B$-field are transformed as 
\begin{align}
 \rmd s'^2 = \Exp{-2a(x)} \rmd y^2 + g_{\mu\nu}(x)\,\rmd x^\mu\,\rmd x^\nu\,,\qquad 
 B'_2 = A_\mu(x)\,\rmd x^\mu \wedge \rmd y\,. 
\end{align}
From $c_y= c$ and \eqref{eq:formula}, we can easily obtain $X'_m$ which, 
together with $G'_{mn}$ and $B'_{mn}$, satisfies the generalized equations;
\begin{align}
 (X'_m) = \begin{pmatrix}
 \partial_\mu (\Phi -a) + c\, A_\mu \\ 
 c\Exp{-2a} 
\end{pmatrix} \,.
\end{align}
This agrees with the known solution (1.10) in \cite{AFHRT}, up to conventions.

\subsubsection*{Solution-generating transformations in mDFT}

As a more general case, let us consider a solution of mDFT of the form,
\begin{align}
 \{\cH_{MN}(x)\,,\ d(x,y)=d_0(x)+c_i\,y^i\,,\ \bX_M(x)\} \,,
\label{eq:solution-ansatz}
\end{align}
as the initial configuration. 
In this case, we can again consider the $\OO(N,N)$ transformations, and if
\begin{align}
 \cH'_{MN}(x) =\Lambda_M{}^K\,\Lambda_N{}^L\,\cH_{KL}(x)\,, \qquad 
 \partial_M d'+ \bX'_M = \Lambda_M{}^N\,\bigl(\partial_N d+ \bX_N\bigr)(x)\,,
\label{eq:soln-generating}
\end{align}
are satisfied, the transformed configuration,
\begin{align}
 \bigl\{\cH'_{MN}(x),\ d'(x,y),\ \bX'_M(x)\bigr\}\,,
\end{align}
is a solution of mDFT. 

\medskip

Here we do not try to find the most general solution of \eqref{eq:soln-generating}, 
but consider a simple case; $\Lie_I B_{mn}=0$\,.
Then, one can take 
\begin{align}
 U_m=-B_{mn}\,I^n \,,
\end{align}
by a suitable redefinition of the dilaton. 
It is further assumed that a coordinate system can be chosen such that the Killing vector $I^m$ becomes constant without violating the ansatz \eqref{eq:solution-ansatz}. 
Then, the second equation in \eqref{eq:soln-generating} is satisfied by
\begin{align}
\begin{split}
 &\Phi'(x,y)=\Phi(x,y) + \ln\Bigl\lvert \frac{\det G'_{mn}}{\det G_{mn}}
 \Bigr\rvert^{1/4} + \bigl[\Lambda_{ij}\,I^j + (\Lambda_i{}^j-\delta_i^j)\,c_j\bigr]\, y^i \,,
\\
 &I'^\mu = I^\mu\,,\qquad I'^i = \Lambda^{ij}\,c_j + \Lambda^i{}_j\,I^j\,,\qquad 
 U'_m = -B'_{mn} \, I'^n \,. 
\end{split}
\label{eq:formula2}
\end{align}
It is easy to show that the conditions in \eqref{eq:conditions} are also satisfied 
in the primed system as long as they are satisfied initially. 
Using this formula, we can generate various solutions of mDFT from a given solution of mDFT. 
Of course, if we set $I^m=0$, the formula \eqref{eq:formula2} is reduced to 
the previous one \eqref{eq:formula}. 

\subsection{Modified F1-charge}
\label{sec:string-charge}

In the conventional supergravity, the equation of motion for the $B$-field is
\begin{align}
 \rmd \bigl(\Exp{-2\Phi} *_{10} H_3\bigr) = 0\,,
\label{eq:eom-B}
\end{align}
where $*_{10}$ is the Hodge star operator associated with the string-frame metric $G_{mn}$. 
If the spacetime has the topology, $\cM_{10}= M_9\times S^1$ with $S^1$ a small circle, 
and closed strings wrapped on $S^1$ are propagating on $M_9$ as point particles, 
the right-hand side of \eqref{eq:eom-B} includes the source terms,
\begin{align}
 \rmd \bigl(\Exp{-2\Phi} *_{10} H_3\bigr) = \sum_p c_p\, \delta^8(x-x_p) \,,
\end{align}
where $x_p$ is the position of the $p$-th particle and $c_p$ is a constant associated 
with the string winding charge. 
Then the total fundamental string charge (F1-charge) can be computed as
\begin{align}
 Q_{\text{F1}} = \int_{V_t} \rmd \bigl(\Exp{-2\Phi} *_{10} H_3\bigr)
 = \int_{\partial V_t} \Exp{-2\Phi} *_{10} H_3 \,,
\end{align}
where $V_t$ is a region on a time-slice of $M_9$ which contains the point particles. 

\medskip

In mDFT, the equations of motion for the $B$-field is $\bs_{[mn]}=0$, namely,
\begin{align}
 \rmd \bigl(\Exp{-2\Phi} *_{10} H_3\bigr) = *_{10} \bigl(\iota_U H_3 + \rmd I \bigr) \,,
\label{eq:B-eom-form}
\end{align}
and apparently we cannot define the F1-charge because there is a cutaneously 
distributed source term on the right-hand side. 

\medskip

In this subsection, we show that the right-hand side of \eqref{eq:B-eom-form} is a closed form,
\begin{align}
 \rmd *_{10} \bigl(\iota_U H_3 + \rmd I \bigr) = 0\,,
\label{eq:closedness}
\end{align}
as far as the null and the generalized Killing properties are satisfied. 
This indicates that, at least locally, there exists a 7-form $\chi_7$, which satisfies
\begin{align}
 \rmd \chi = *_{10} \bigl(\iota_U H_3 + \rmd I \bigr) \,, 
\end{align}
and we can define a modified 7-form field strength,
\begin{align}
 H'_7 \equiv \Exp{-2\Phi} *_{10} H_3 - \chi_7 \,,
\end{align}
satisfying the Bianchi identity, $\rmd H'_7 = 0$. 
This allows to define a modified string charge as
\begin{align}
 Q'_{\text{F1}} \equiv \int_{V_t} \rmd H'_7 = \int_{\partial V_t} H'_7 \,. 
\end{align}

\medskip

In order to show the closedness \eqref{eq:closedness}, let us rewrite \eqref{eq:B-eom-form} 
in terms of the components as
\begin{align}
 \frac{1}{2}\,\partial_k \bigl(\Exp{-2\Phi}\sqrt{\abs{G}} H^{kmn}\bigr) 
 = \Exp{-2\Phi}\sqrt{\abs{G}}\, \bigl(U_k\,H^{kmn} + D^m I^n - D^n I^m \bigr) \,. 
\label{eq:closedness-components}
\end{align}
Then the closedness means the divergence-free condition of the right-hand side of \eqref{eq:closedness-components},
\begin{align}
 \partial_m \bigl[\Exp{-2\Phi}\sqrt{\abs{G}}\, \bigl(U_k\,H^{kmn} + D^m I^n - D^n I^m \bigr)\bigr] = 0\,,
\end{align}
or equivalently,
\begin{align}
 D_m \bigl[\Exp{-2\Phi} \bigl(U_k\,H^{kmn} + D^m I^n - D^n I^m \bigr)\bigr] = 0\,. 
\end{align}
A straightforward calculation shows
\begin{align}
 &\Exp{2\Phi}D_m \bigl[\Exp{-2\Phi} \bigl(U^k\,H_k{}^{mn} + D^m I^n - D^n I^m \bigr)\bigr]
\nn\\
 &= -2\,\bigl(\bs^{[nk]}\,U_k + \bs^{(nk)}\,I_k \bigr)
   + \tfrac{1}{2}\,\bigl(H^{nlk} + 4\,G^{nl} I^k \bigr)\,\bigl(I^m\,H_{mkl} + D_k U_l - D_l U_k\bigr)
\nn\\
 &\quad +2 D^n \bigl(2\,I^k\,Z_k - D_m I^m \bigr) 
        + \bigl(D_k -2\,Z_k \bigr) \bigl(D^k I^n + D^n I^k\bigr) \,.
\end{align}
Hence, as long as the null and the generalized Killing properties are satisfied, the closedness condition 
\eqref{eq:closedness} is satisfied under the equations of motion and $\chi_7$ can be found at least locally. 

\medskip

Given a solution of mDFT or the generalized supergravity, it should satisfy $\rmd H'_7=0$ and 
there exists the associated potential $B'_6$ satisfying 
\begin{align}
 H'_7 = \rmd B'_6 \,. 
\end{align}
In the absence of the extra generalized vector $\bX^M$, this potential is reduced 
to the conventional 6-form $B_6$, which minimally couples to the NS5-brane. 
Thus it is quite natural to expect that the NS5-brane propagating in a background, 
which satisfies the equations of motion of mDFT, will minimally couples to the modified potential $B'_6$\,. 
It would be an interesting future problem to construct a world-volume action 
of the NS5-brane propagating in backgrounds of mDFT.

\subsection{An attempt to construct the classical action}
\label{sec:action}

In DFT, the generalized Ricci flatness conditions are derived as the equations of motion. 

\medskip

In this subsection, let us first recall how to derive the flatness condition of 
the generalized Ricci tensors. 
We will then discuss why it is so difficult to realize 
the modified generalized Ricci flatness conditions from the action principle. 

\medskip

Let us consider the Lagrangian density for DFT,
\begin{align}
 \cL =\Exp{-2d}\,\cS = \Exp{-2d}\,\bigl(P^{MK}\,P^{NL}
 -\bar{P}^{MK}\,\bar{P}^{NL}\bigr)\, S_{MNKL} \,.
\end{align}
By taking a variation, the following expression is obtained \cite{Park:2015bza}:
\begin{align}
 \delta \cL &=-2 \Exp{-2d} \cS\,\delta d - \tfrac{1}{2} \Exp{-2d}\cS_{MN}\, \delta \cH^{MN} 
 + \Exp{-2d}\,\nabla_M \bigl[2\,\bigl(P^{MK}\,P^{NL}
 -\bar{P}^{MK}\,\bar{P}^{NL}\bigr)\, \delta\Gamma_{NKL} \bigr]
\nn\\
 &= - \Exp{-2d} \bigl(2 \cS\,\delta d + \tfrac{1}{2} \,\cS_{MN}\, \delta \cH^{MN}\bigr) 
 + \Exp{-2d} \nabla_M \bigl(4\,\cH^{MN}\,\partial_N \delta d 
 - \nabla_N\,\delta \cH^{MN}\bigr) \,.
\end{align}
Thanks to the volume factor $\Exp{-2d}$ and \eqref{eq:total-div}, 
the last term can be rewritten as
\begin{align}
 \delta \cL = - \Exp{-2d} \bigl(2\,\cS\,\delta d + \tfrac{1}{2}\,\cS_{MN}\, 
 \delta \cH^{MN}\bigr) + \partial_M \bigl[\Exp{-2d}\bigl(4\,\cH^{MN}\,
 \partial_N \delta d - \nabla_N\,\delta \cH^{MN}\bigr)\bigr] \,. 
\end{align}
Note here that the total derivative terms do not contribute to the equations of motion. 
Then the action principle leads to the generalized Ricci flatness conditions, 
$\cS_{MN}=0$ and $\cS=0$\,. 
That is, the equations of motion are described as the flatness condition of the generalized Ricci tensors.

\medskip 

Instead, let us try to consider a modified Lagrangian density,
\begin{align}
 \cL' =\Exp{-2d} \bS = \Exp{-2d}\,
       \bigl(P^{MK}\,P^{NL}-\bar{P}^{MK}\,\bar{P}^{NL}\bigr)\,\bS_{MNKL} \,. 
\end{align} 
Its variation with respect to $\cH_{MN}(x)$ and $d(x)$ gives
\begin{align}
 \delta \cL' &=\Exp{-2d} \Bigl[-2\,\bS\,\delta d - \tfrac{1}{2}\,\bS_{MN}\, \delta \cH^{MN} 
 + \bnabla_M \bigl(4\,\cH^{MN}\,\partial_N \delta d - \bnabla_N\,\delta \cH^{MN}\bigr)\Bigr]\,,
\end{align}
and from \eqref{eq:div}, we obtain
\begin{align}
 \delta \cL' &=-\Exp{-2d} \bigl(2\,\bS\,\delta d + \tfrac{1}{2}\,\bS_{MN}\, 
 \delta \cH^{MN} \bigr)
 + \partial_M \bigl[\Exp{-2d}\bigl(4\,\cH^{MN}\,\partial_N \delta d 
 - \bnabla_N\,\delta \cH^{MN}\bigr)\bigr] 
\nn\\
 &\quad - 2\Exp{-2d}\bX_M\,\bigl(4\,\cH^{MN}\,\partial_N \delta d 
 - \bnabla_N\,\delta \cH^{MN}\bigr) 
\nn\\
 &= -2\Exp{-2d} \bigl(\bS - 4\,\cH^{MN}\,\nabla_M\bX_N\bigr)\,\delta d 
 - \tfrac{1}{2} \Exp{-2d} \bigl(\bS_{MN}+4\,\nabla_{(M} \bX_{N)}
 +8\,\bX_M \bX_N\bigr)\, \delta \cH^{MN} 
\nn\\
 &\quad + \partial_M \bigl[\Exp{-2d}\bigl(4\,\cH^{MN}\,\partial_N \delta d 
 - 8\,\bX_N\,\cH^{MN}\,\delta d - \bnabla_N\,\delta \cH^{MN}+2\,\bX_N\,
 \delta\cH^{MN}\bigr)\bigr] \,.
\end{align}
Then, the equations of motion become%
\footnote{The same result can be easily obtained if we note that $\Exp{-2d} \bS$ 
is equal to $\Exp{-2d} \bigl(\cS - 4\,\cH_{MN}\,\bX^M \bX^N\bigr)$ 
up to a total derivative term.}
\begin{align}
 &\bS - 4\,\cH^{MN}\,\nabla_M\bX_N = 0\,,
\nn\\
 &\bS_{MN}+4\,\bigl(P_M{}^K\,\bar{P}_N{}^L + \bar{P}_M{}^K\,P_N{}^L\bigr)\,
 \bigl(\nabla_{(K} \bX_{L)}+ 2\,\bX_K \bX_L\bigr) = 0 \,,
\end{align}
or equivalently,
\begin{align}
 \cS = 4\,\cH^{MN}\,\bX_M\bX_N \,,\qquad 
 \cS_{MN}= -8\,\bigl(P_M{}^K\,\bar{P}_N{}^L + \bar{P}_M{}^K\,P_N{}^L\bigr)\, \bX_K \bX_L \,. 
\end{align}
These are not of our interest and $\cL'$ is not a correct Lagrangian density for mDFT. 

\medskip 

A possible way is to find out a new volume factor $\omega$ which satisfies
\begin{align}
 \omega\,\bnabla_M V^M = \partial_M \bigl(\omega\, V^M\bigr) \,, 
\label{eq:omega-condition}
\end{align}
for an arbitrary generalized vector $V^M$ and changes under the variation as
\begin{align}
 \delta \omega ~~\propto~~ \omega\, \delta d + \omega\,\alpha^{MN}\, \delta \cH_{MN} \,. 
\end{align}
Here, $\alpha^{MN}$ is a certain generalized tensor that vanishes when $\bS=0$ is satisfied. 
If we could find such $\omega$\,, the Lagrangian density,
\begin{align}
 \cL^\bX = \omega\, \bS \,,
\end{align}
gives the modified generalized Ricci flatness conditions. 
Indeed, the variation becomes 
\begin{align}
 \delta \cL^\bX 
 = \bS\,\delta \omega - \tfrac{1}{2}\,\omega\,\bS_{MN}\, \delta \cH^{MN} 
 + \partial_M \bigl[\omega\,\bigl(4\,\cH^{MN}\,\partial_N \delta d 
- \bnabla_N\,\delta \cH^{MN}\bigr)\bigr] \,,
\end{align}
and the desired relations $\bS_{MN}=0$ and $\bS=0$ are obtained. 
However, \eqref{eq:omega-condition} is equivalent to
\begin{align}
 \bGamma_M = \partial_M \ln \omega \,,
\end{align}
and recalling $\bGamma_M = -2\,\bigl(\partial_M d + \bX_M\bigr)$\,, 
the generalized vector $\bX_M$ must be of the form,
\begin{align}
 \bX_M = \partial_M f\,. 
\end{align}
It is so restrictive that $\bX_M$ can be removed by a redefinition of the dilaton. 
As a result, the system goes back to the conventional DFT. 
Thus, it is difficult to find a good Lagrangian density for mDFT. 
It still remains to be solved. 

\medskip 

As a closely related issue, let us make a brief comment on the differential Bianchi identity \eqref{eq:diff-Bianch}. 
In DFT, this can be derived from a invariance of the action under an infinitesimal generalized diffeomorphism \cite{Siegel:1993th,Kwak:2010ew,Hohm:2010xe}. 
Unfortunately, since we do not have the action of mDFT, we cannot derive the modified differential Bianchi identity, like
\begin{align}
 \bnabla^M \bS ~ \sim ~ 2\,\cH^{MK}\,\bnabla^N \bS_{NK}\,,
\end{align}
from a similar consideration. 
However, if we choose the canonical section $(\partial_M)=(\partial_m,\,0)$ 
and adopt the parameterizations \eqref{eq:cH-parametrization}, 
we can show a similar identity,
\begin{align}
 D^k \bS 
 &= 2\,\bigl(D_l-2\,Z_l\bigr)\, \bs^{(lk)} - \bigl(H^{kpq}+4\,G^{kp}\, I^q\bigr)\, \bs_{[pq]} 
   + 2\,\bigl(D_l-2\,Z_l\bigr)\,\bigl(I_p\, H^{pkl} + D^k Z^l - D^l Z^k\bigr)
\nn\\
  &\quad - 4\, \bigl[ I^l\,\bigl(D_k I_l + D_l I_k\bigr) + D_k \bigl(2\,I^l\,Z_l-D_l I^l\bigr) \bigr] \,.
\end{align}
In particular, if we use the null and the generalized Killing properties, this becomes
\begin{align}
 D^k \bS 
 = 2\,\bigl(D_l-2\,Z_l\bigr)\, \bs^{(lk)} - \bigl(H^{kpq}+4\,G^{kp}\, I^q\bigr)\, \bs_{[pq]} \,. 
\end{align}
From this identity, we can show that $\bS_{NM}=0$, or equivalently $\bs_{mn}=0$, leads to $\partial_k \bS=0$ as it was shown in \cite{AFHRT}. 
It will be interesting to investigate a $\OO(D,D)$ covariant expression of this Bianchi identity without choosing the canonical section. 

\section{Conclusion and Outlook}
\label{sec:conclusion}

In this paper, we have considered a modified generalized covariant derivative $\bnabla_M$ 
in a doubled spacetime, relaxing a condition of the covariant constancy of 
the DFT dilaton, $\nabla_M d = 0$, as
\begin{align}
 \bnabla_M d = -\bX_M \,,
\end{align}
with an extra generalized vector $\bX_M$\,. 
Then we have studied a modification of the equations of 
motion in DFT, $\bS_{MN}=0$ and $\bS=0$\,, and shown that these reproduces the NS-NS part of the generalized equations of type IIB supergravity if the generalized vector $\bX_M$ is assumed to have the null and the generalized Killing properties. 
We have also studied the global transformations 
which map a solution of mDFT to other solutions of mDFT, generalizing the known generalized 
$T$-duality transformations (for the NS-NS fields) in the generalized supergravity. 
A subtle issue in defining the F1-charge in mDFT has also been discussed, 
and a definition of the modified F1-charge is proposed. 

\medskip 

One of the most important issues to be resolved is the construction of the mDFT action. 
As we have discussed in section \ref{sec:action}, the action allows us to derive the modified differential Bianchi identity. 
Also, it allows us to construct Noether currents associated with certain symmetries. 
For example, in \cite{Blair:2015eba,Park:2015bza}, Noether currents associated with doubled spacetime isometries are studied from the DFT action, and expressions for the ADM momenta and the F1-charges are obtained. 
If we could perform a similar analysis in mDFT, it would be possible to derive the modified F1-charges as the Noether currents. 
The action also allows to discuss the black hole thermodynamics by following Wald's approach \cite{Wald:1993nt,Iyer:1994ys,Iyer:1995kg} 
(for the black hole thermodynamics in DFT, see \cite{Arvanitakis:2016zes}). 
In mDFT, because of a modification made in the dilaton sector, the Einstein frame metric and accordingly the Bekenstein-Hawking entropy may be modified as well. 
As discussed in section \ref{sec:action}, to construct the action, the construction of a suitable volume form will be important. 

\medskip

%

As a generalization of mDFT, it is important to include the R-R fields as well. 
One approach is to extend the formulation developed in \cite{Hohm:2011zr,Hohm:2011dv,Jeon:2012kd} by introducing the same generalized vector $\bX^M$. 
However, this approach might not be satisfactory in the sense that it treats the NS-NS sector 
and the R-R sector differently; the NS-NS fields are contained in the generalized metric 
but the R-R fields are treated as matters. 

\medskip

Another approach is a $U$-duality covariant formulation, 
called the exceptional field theory (EFT) \cite{West:2001as,West:2003fc,West:2004st,Hull:2007zu,Pacheco:2008ps,Hillmann:2009ci,Berman:2010is,Berman:2011pe,Berman:2011cg,Berman:2011jh,Berman:2012vc,Park:2013gaj,Aldazabal:2013mya,Hohm:2013pua,Hohm:2013vpa,Hohm:2013uia,Godazgar:2014nqa,Hohm:2014fxa,Musaev:2014lna,Hohm:2015xna,Abzalov:2015ega,Musaev:2015ces,Berman:2015rcc,Ciceri:2016dmd,Baguet:2016jph}. 
In $E_{d(d)}$ EFT, all of the bosonic fields in the 11D supergravity can be packaged into the generalized metric and certain tensors that depend on the dimension $d$. 
After choosing a suitable section, the EFT action reproduces the bosonic part of the conventional 11D supergravity. 
Interestingly, we can also consider a parameterization in terms of the bosonic fields in 10D type IIB supergravity and the same EFT action can reproduce the bosonic part of type IIB supergravity action as well. 
The construction of the generalized Ricci tensors in EFT is studied in \cite{Aldazabal:2013mya} (see also \cite{Coimbra:2012af}) and the equations of motion in the 11D supergravity and the type IIB supergravity are expressed by using the generalized Ricci tensor. 
Then, it is interesting to consider a modification of the generalized connection, as we have done in this paper (i.e., modification of the condition (4.12) in \cite{Aldazabal:2013mya}), and see whether the generalized equations of type IIB supergravity are reproduced just by replacing the generalized Ricci tensor with the modified one. 

\medskip 

For ensuring the above direction, the $U$-duality covariance is necessary to be realized in the generalized equations of type IIB supergravity. 
Hence it will be an important task to study the $S$-duality covariance of the generalized equations. 

\medskip 

We hope that the proposed mDFT would shed light on unexplored physics hiding 
behind the generalized supergravity.

\subsection*{Addendum}

After this paper has appeared on the arXiv, Baguet, Magro and Samtleben submitted 
an interesting paper \cite{Baguet:2016prz}, where the equations of the generalized type IIB 
supergravity are derived from the exceptional field theory 
by choosing a section condition for the type IIA description with a Scherk-Schwarz ansatz. 
It is worth noting that the same derivation (which is restricted to the NS-NS sector though) 
has already been explained in section \ref{sec:DFT-mDFT} of this paper, 
without using the term of the Scherk-Schwarz reduction. 

\medskip 

In sections \ref{sec:modified-Ricci} and \ref{sec:EOM-mDFT}, we have explained that 
the generalized equations can be expressed in a manifestly covariant way 
(i.e.~the flatness condition for the modified generalized Ricci tensors). 
There, an extra generalized vector $\bX^M$\,, which is not contained in the conventional DFT, 
has been introduced as a modification in the generalized connection. 
On the other hand, in section \ref{sec:DFT-mDFT}, we have constructed solutions of 
the generalized equations with constant $\bX^M$ by performing the $\OO(N,N)$ rotations 
to a solution of the conventional DFT. 
In fact, the latter procedure is essentially the same as the Scherk-Schwarz reduction 
considered in \cite{Baguet:2016prz}, and $\partial_M d'+\bX'_M$ introduced in \eqref{eq:DFT-GS} 
is precisely a combination $\partial_M \hat{d} + f_M$ defined below. 
Namely, in section \ref{sec:DFT-mDFT}, we have introduced the additional generalized vector $\bX_M$ 
as the gauging $f_M$, in terms of the Scherk-Schwarz reduction. 

\medskip

The Scherk-Schwarz reduction in DFT has been studied in \cite{Aldazabal:2011nj,Geissbuhler:2011mx,Grana:2012rr}. 
In terms of the generalized metric, the ansatz is given by
\begin{align}
 \cH_{MN}(x,y) = (\Lambda^{\rmT})_M{}^K(y)\, \hat{\cH}_{KL}(x)\, \Lambda^L{}_N(y)\,,\quad 
 d(x,y) = \hat{d}(x) + \lambda(y) \,. 
\label{eq:Scherk-Schwarz}
\end{align}
The generalized Ricci scalar $\cS$ for $\{\cH_{MN}(x,y),\,d(x,y)\}$ has been computed in \cite{Aldazabal:2011nj,Geissbuhler:2011mx,Grana:2012rr}. 
In a special case where $\Lambda^I{}_J$ is constant and $\lambda(y)=c_i\,y^i$, 
which corresponds to our ansatz \eqref{eq:linear-solution} rotated by the $\OO(N,N)$ transformation \eqref{eq:ONN}, 
$\cS$ becomes a sum of $\cS$ for $\{\hat{\cH}_{MN}(x),\,\hat{d}(x)\}$ and 
an additional term $\cS_f$, 
\begin{align}
 \cS_f \equiv -2\,f_M\,\partial_N \hat{\cH}^{MN} +4\,f_M\, \hat{\cH}^{MN}\,\partial_N \hat{d} - f_M\,f_N\, \hat{\cH}^{MN} \,, \quad 
 f_M \equiv -2\,(\Lambda^{-\rmT})_M{}^N\,\partial_N \lambda \,. 
\end{align}
According to $\partial_M f_N=0$ and an identification $f_M= -2\,\bX_M$, 
$\cS_f$ agrees with the modification $\delta_\bX \cS$ in \eqref{eq:delta-X-cS}. 
The modification of the generalized Ricci curvature $\delta_\bX \cS_{MN}$ can also be 
reproduced in the same manner. 
Furthermore, the requirements for the gauging \cite{Aldazabal:2011nj,Geissbuhler:2011mx,Grana:2012rr},
\begin{align}
 f^M\,f_M =0 \,,\quad f^M\,\partial_M \hat{\cH}_{KL}(x)=0\,,\quad f^M\,\partial_M \hat{d}(x)=0\,,
\end{align}
which are indeed satisfied by
\begin{align}
 f_M = -2\,(\Lambda^{-\rmT})_M{}^N\,\partial_N \lambda = -2\,\bigl(0\,,\ \Lambda_i{}^j\,c_j \,,\ 0 \,, \ \Lambda^{ij}\,c_j\bigr)^\rmT \,,
\end{align}
correspond to the null and the generalized Killing properties for $\bX^M$. 
Note that a combination $\partial_M \hat{d} + f_M$ is appearing as $\partial_M d'+\bX'_M$ in \eqref{eq:DFT-GS}.\footnote{Here, we made an identification $\{d',\,\bX'_M\}=\{\hat{d},\,-f_M/2\}$. On the other hand, in \eqref{eq:sol-mDFT}, the decomposition of $\partial_M d'+\bX'_M$ into $\{d',\,\bX'_M\}$ is made so that $(\bX'_M)$ should take the form, $(\bX'_M)=(0,\,I'^m)$. For the configuration \eqref{eq:sol-DFT} which corresponds to 
$\Lambda^I{}_J=\delta^I_J$, the same decomposition leads to $\bX_M=0$ and 
it is a solution of the conventional DFT.}

\medskip 

In addition, as it was shown in \cite{Hohm:2014qga}, when we consider a Scherk-Schwarz reduction in EFT, the action principle does not work for the low-dimensional theory if there is a non-vanishing trombone gauging $\vartheta_M$. 
The trombone gauging $\vartheta_M$ appears to be identified with the above introduced gauging $f_M$ in DFT, or our extra generalized vector $\bX_M$. 
If this identification is correct, the difficulty in the construction of the action discussed in section \ref{sec:action} has the same origin as the EFT case discussed in \cite{Hohm:2014qga}.\footnote{We would like to thank the anonymous referee for informing us of the interesting paper \cite{Hohm:2014qga}.} 
It will be interesting to clarify this matter in more detail. 

\subsection*{Acknowledgments}

K.Y.\ is very grateful to J.~Sakamoto for useful discussions. 
K.Y.\ would like to appreciate useful discussions during the workshop 
``Generalized Geometry \& T-dualities'' at the Simons center for geometry and physics, 
especially inspiring comments from A.~A.~Tseytlin and C.~Klimcik. 
The work of K.Y.\ was supported by the Supporting Program for Interaction-based 
Initiative Team Studies (SPIRITS) from Kyoto University and 
by a JSPS Grant-in-Aid for Scientific Research (C) No.\,15K05051.
This work was also supported in part by the JSPS Japan-Russia Research 
Cooperative Program and the JSPS Japan-Hungary Research Cooperative Program.


\begin{thebibliography}{99}
\bibitem{WT} 
A.~A.~Tseytlin and  L.~Wulff,
  ``Kappa-symmetry of superstring sigma model and generalized 10d supergravity equations,'' 
JHEP {\bf 1606} (2016) 174
  [arXiv:1605.04884 [hep-th]].

\bibitem{AFHRT}  
G.~Arutyunov, S.~Frolov, B.~Hoare, R.~Roiban and A.~A.~Tseytlin,
  ``Scale invariance of the eta-deformed AdS$_5\times$S$^5$ superstring, 
T-duality and modified type II equations,'' 
Nucl.\ Phys.\ B {\bf 903} (2016) 262
  [arXiv:1511.05795 [hep-th]].

\bibitem{Townsend}
M.~T.~Grisaru, P.~Howe, L.~Mezincescu, B.~E.~W.~Nilsson and P.~K.~Townsend, 
``$N=2$ superstrings in a supergravity background,'' 
Phys.\ Lett.\ B {\bf 162} (1985) 116-120. \\ 
E.~Bergshpeff, E.~Sezgin and P.~K.~Townsend, 
``Superstring actions in $D=3,4,6,10$ curved superspace,'' 
Phys.\ Lett.\ B {\bf 169} (1986) 191-196. 

\bibitem{K}
 C.~Klimcik,
  ``Yang-Baxter sigma models and dS/AdS T duality,''  
JHEP {\bf 0212} (2002) 051  [hep-th/0210095]; 
  ``On integrability of the Yang-Baxter sigma-model,''  
J.\ Math.\ Phys.\  {\bf 50} (2009) 043508  [arXiv:0802.3518 [hep-th]]; 
 ``Integrability of the bi-Yang-Baxter sigma model,''  Lett.\ Math.\ Phys.\  {\bf 104} (2014) 1095
  [arXiv:1402.2105 [math-ph]].  

\bibitem{DMV1}
  F.~Delduc, M.~Magro and B.~Vicedo,
  ``On classical q-deformations of integrable sigma-models,''  
  JHEP {\bf 1311} (2013) 192  [arXiv:1308.3581 [hep-th]].   
  
\bibitem{MY}
 T.~Matsumoto and K.~Yoshida,
  ``Yang-Baxter sigma models based on the CYBE,''
  Nucl.\ Phys.\ B {\bf 893} (2015) 287
  [arXiv:1501.03665 [hep-th]].   

\bibitem{DMV2}
  F.~Delduc, M.~Magro and B.~Vicedo,
  ``An integrable deformation of the AdS$_5\times$S$^5$ superstring action,''  
 Phys.\ Rev.\ Lett.\  {\bf 112} (2014) 051601
  [arXiv:1309.5850 [hep-th]];   
  ``Derivation of the action and symmetries of the $q$-deformed AdS$_5\times$S$^5$ superstring,'' 
  JHEP {\bf 1410} (2014) 132
  [arXiv:1406.6286 [hep-th]].  
  
\bibitem{KMY}
  I.~Kawaguchi, T.~Matsumoto and K.~Yoshida,
  ``Jordanian deformations of the AdS$_5\times$S$^5$ superstring,''
  JHEP {\bf 1404} (2014) 153
  [arXiv:1401.4855 [hep-th]].  

\bibitem{DJ}
  V.~G.~Drinfel'd,
  ``Hopf algebras and the quantum Yang-Baxter equation,'' 
  Sov.\ Math.\ Dokl.\ {\bf 32} (1985) 254;``Quantum groups,''
  J.\ Sov.\ Math.\  {\bf 41} (1988) 898 
  [Zap.\ Nauchn.\ Semin.\  {\bf 155}, 18 (1986)]. 
  M.~Jimbo,
  ``A $q$ difference analog of $U(g)$ and the Yang-Baxter equation,''
  Lett.\ Math.\ Phys.\  {\bf 10} (1985) 63.      

\bibitem{ABF}
    G.~Arutyunov, R.~Borsato and S.~Frolov,
  ``S-matrix for strings on $\eta$-deformed AdS$_5\times$S$^5$\,s,'' 
  JHEP {\bf 1404} (2014) 002 [arXiv:1312.3542 [hep-th]].  

\bibitem{ABF2}
G.~Arutyunov, R.~Borsato and S.~Frolov,
  ``Puzzles of eta-deformed AdS$_5\times$S$^5$\,,'' 
JHEP {\bf 1512} (2015) 049
  [arXiv:1507.04239 [hep-th]]. 

\bibitem{HT-sol}
  B.~Hoare and A.~A.~Tseytlin,
  ``Type IIB supergravity solution for the T-dual of the $\eta$-deformed AdS$_{5}\times$S$^{5}$ superstring,''
  JHEP {\bf 1510} (2015) 060 [arXiv:1508.01150 [hep-th]].

\bibitem{HT}
B.~Hoare and A.~A.~Tseytlin,
  ``On integrable deformations of superstring sigma models related to AdS$_n \times$S$^n$ 
supercosets,''
  Nucl.\ Phys.\ B {\bf 897} (2015) 448
  [arXiv:1504.07213 [hep-th]]. 


\bibitem{LM-MY}
  T.~Matsumoto and K.~Yoshida,
  ``Lunin-Maldacena backgrounds from the classical Yang-Baxter equation 
- towards the gravity/CYBE correspondence,''
  JHEP {\bf 1406} (2014) 135
  [arXiv:1404.1838 [hep-th]].    

\bibitem{MR-MY}  
 T.~Matsumoto and K.~Yoshida,
  ``Integrability of classical strings dual for noncommutative gauge theories,''
  JHEP {\bf 1406} (2014) 163 
  [arXiv:1404.3657 [hep-th]].  
  
\bibitem{Sch-MY}
  T.~Matsumoto and K.~Yoshida,
  ``Schr\"odinger geometries arising from Yang-Baxter deformations,'' 
  JHEP {\bf 1504} (2015) 180 [arXiv:1502.00740 [hep-th]].  
  
\bibitem{KMY-SUGRA}
I.~Kawaguchi, T.~Matsumoto and K.~Yoshida,
 ``A Jordanian deformation of AdS space in type IIB supergravity,'' 
 JHEP {\bf 1406} (2014) 146 [arXiv:1402.6147 [hep-th]].        
       
\bibitem{MY-duality}
  T.~Matsumoto and K.~Yoshida,
  ``Yang-Baxter deformations and string dualities,''
  JHEP {\bf 1503} (2015) 137 [arXiv:1412.3658 [hep-th]]. 

\bibitem{MY-summary}
  T.~Matsumoto and K.~Yoshida,
  ``Integrable deformations of the AdS$_{5} \times S^5$ superstring 
and the classical Yang-Baxter equation 
{\it -Towards the gravity/CYBE correspondence-},''
  J.\ Phys.\ Conf.\ Ser.\  {\bf 563} (2014) 1,  012020
  [arXiv:1410.0575 [hep-th]]; 
   ``Towards the gravity/CYBE correspondence -- current status --,''
  J.\ Phys.\ Conf.\ Ser.\  {\bf 670} (2016) no.1,  012033. 

\bibitem{Stijn1}
S.~J.~van Tongeren,
 ``On classical Yang-Baxter based deformations of the AdS$_5 \times$S$^5$ superstring,''
 JHEP {\bf 1506} (2015) 048 [arXiv:1504.05516 [hep-th]]. 
 
\bibitem{Stijn2}
S.~J.~van Tongeren, 
 ``Yang-Baxter deformations, AdS/CFT, and twist-noncommutative gauge theory,'' 
 Nucl.\ Phys.\ B {\bf 904} (2016) 148 
 [arXiv:1506.01023 [hep-th]].
  
\bibitem{CMY}
   P.~M.~Crichigno, T.~Matsumoto and K.~Yoshida,
   ``Deformations of $T^{1,1}$ as Yang-Baxter sigma models,''
    JHEP {\bf 1412} (2014) 085
   [arXiv:1406.2249 [hep-th]]; 
    ``Towards the gravity/CYBE correspondence beyond integrability 
 -- Yang-Baxter deformations of $T^{1,1}$,'' 
  J.\ Phys.\ Conf.\ Ser.\  {\bf 670} (2016) no.1,  012019 
   [arXiv:1510.00835 [hep-th]].

\bibitem{KY} 
  H.~Kyono and K.~Yoshida,
  ``Supercoset construction of Yang-Baxter deformed AdS$_5\times$S$^5$ backgrounds,''
Prog.\ Theor.\ Exp.\ Phys.\ (2016) 083B03 
  [arXiv:1605.02519 [hep-th]].

\bibitem{HvT} 
  B.~Hoare and S.~J.~van Tongeren,
  ``On jordanian deformations of AdS$_5$ and supergravity,''
 J.\ Phys.\ A {\bf 49} (2016) no.43,  434006
  [arXiv:1605.03554 [hep-th]].

\bibitem{ORSY}
D.~Orlando, S.~Reffert, J.~i.~Sakamoto and K.~Yoshida,
  ``Generalized type IIB supergravity equations and non-Abelian classical r-matrices,''
  J.\ Phys.\ A {\bf 49} (2016) no.44,  445403
  [arXiv:1607.00795 [hep-th]].

\bibitem{BW}
  R.~Borsato and L.~Wulff,
  ``Target space supergeometry of $\eta$ and $\lambda$-deformed strings,''
  JHEP {\bf 1610} (2016) 045
  [arXiv:1608.03570 [hep-th]].

\bibitem{OvT}
  D.~Osten and S.~J.~van Tongeren,
  ``Abelian Yang-Baxter Deformations and TsT transformations,''
  arXiv:1608.08504 [hep-th].

\bibitem{Stijn3}
  S.~J.~van Tongeren,
  ``Almost abelian twists and AdS/CFT,''
  arXiv:1610.05677 [hep-th].

\bibitem{Siegel:1993xq}
  W.~Siegel,
  ``Two vierbein formalism for string inspired axionic gravity,''
  Phys.\ Rev.\ D {\bf 47} (1993) 5453
  [hep-th/9302036].



\bibitem{Siegel:1993th}
  W.~Siegel,
  ``Superspace duality in low-energy superstrings,''
  Phys.\ Rev.\ D {\bf 48} (1993) 2826
  [hep-th/9305073].



\bibitem{Siegel:1993bj}
  W.~Siegel,
  ``Manifest duality in low-energy superstrings,''
  hep-th/9308133.



\bibitem{Hull:2009mi}
  C.~Hull and B.~Zwiebach,
  ``Double Field Theory,''
  JHEP {\bf 0909} (2009) 099
  [arXiv:0904.4664 [hep-th]].



\bibitem{Hull:2009zb}
  C.~Hull and B.~Zwiebach,
  ``The Gauge algebra of double field theory and Courant brackets,''
  JHEP {\bf 0909} (2009) 090
  [arXiv:0908.1792 [hep-th]].



\bibitem{Hohm:2010jy}
  O.~Hohm, C.~Hull and B.~Zwiebach,
  ``Background independent action for double field theory,''
  JHEP {\bf 1007} (2010) 016
  [arXiv:1003.5027 [hep-th]].



\bibitem{Hohm:2010pp}
  O.~Hohm, C.~Hull and B.~Zwiebach,
  ``Generalized metric formulation of double field theory,''
  JHEP {\bf 1008} (2010) 008
  [arXiv:1006.4823 [hep-th]].



\bibitem{Jeon:2010rw}
  I.~Jeon, K.~Lee and J.~H.~Park,
  ``Differential geometry with a projection: Application to double field theory,''
  JHEP {\bf 1104} (2011) 014
  [arXiv:1011.1324 [hep-th]].



\bibitem{Hohm:2010xe}
  O.~Hohm and S.~K.~Kwak,
  ``Frame-like Geometry of Double Field Theory,''
  J.\ Phys.\ A {\bf 44} (2011) 085404
  [arXiv:1011.4101 [hep-th]].



\bibitem{Jeon:2011cn}
  I.~Jeon, K.~Lee and J.~H.~Park,
  ``Stringy differential geometry, beyond Riemann,''
  Phys.\ Rev.\ D {\bf 84} (2011) 044022
  [arXiv:1105.6294 [hep-th]].



\bibitem{Hohm:2011zr}
  O.~Hohm, S.~K.~Kwak and B.~Zwiebach,
  ``Unification of Type II Strings and T-duality,''
  Phys.\ Rev.\ Lett.\  {\bf 107} (2011) 171603
  [arXiv:1106.5452 [hep-th]].



\bibitem{Hohm:2011dv}
  O.~Hohm, S.~K.~Kwak and B.~Zwiebach,
  ``Double Field Theory of Type II Strings,''
  JHEP {\bf 1109} (2011) 013
  [arXiv:1107.0008 [hep-th]].



\bibitem{Jeon:2011vx}
  I.~Jeon, K.~Lee and J.~H.~Park,
  ``Incorporation of fermions into double field theory,''
  JHEP {\bf 1111} (2011) 025
  [arXiv:1109.2035 [hep-th]].



\bibitem{Hohm:2011si}
  O.~Hohm and B.~Zwiebach,
  ``On the Riemann Tensor in Double Field Theory,''
  JHEP {\bf 1205} (2012) 126
  [arXiv:1112.5296 [hep-th]].



\bibitem{Hohm:2011nu}
  O.~Hohm and S.~K.~Kwak,
  ``N=1 Supersymmetric Double Field Theory,''
  JHEP {\bf 1203} (2012) 080
  [arXiv:1111.7293 [hep-th]].



\bibitem{Jeon:2011sq}
  I.~Jeon, K.~Lee and J.~H.~Park,
  ``Supersymmetric Double Field Theory: Stringy Reformulation of Supergravity,''
  Phys.\ Rev.\ D {\bf 85} (2012) 081501
   Erratum: [Phys.\ Rev.\ D {\bf 86} (2012) 089903]
  [arXiv:1112.0069 [hep-th]].



\bibitem{Jeon:2012kd}
  I.~Jeon, K.~Lee and J.~H.~Park,
  ``Ramond-Ramond Cohomology and O(D,D) T-duality,''
  JHEP {\bf 1209} (2012) 079
  [arXiv:1206.3478 [hep-th]].



\bibitem{Jeon:2012hp}
  I.~Jeon, K.~Lee, J.~H.~Park and Y.~Suh,
  ``Stringy Unification of Type IIA and IIB Supergravities under N=2 D=10 Supersymmetric Double Field Theory,''
  Phys.\ Lett.\ B {\bf 723} (2013) 245
  [arXiv:1210.5078 [hep-th]].



\bibitem{Zwiebach:2011rg}
  B.~Zwiebach,
  ``Double Field Theory, T-Duality, and Courant Brackets,''
  Lect.\ Notes Phys.\  {\bf 851} (2012) 265
  [arXiv:1109.1782 [hep-th]].



\bibitem{Aldazabal:2013sca}
  G.~Aldazabal, D.~Marques and C.~Nunez,
  ``Double Field Theory: A Pedagogical Review,''
  Class.\ Quant.\ Grav.\  {\bf 30} (2013) 163001
  [arXiv:1305.1907 [hep-th]].



\bibitem{Berman:2013eva}
  D.~S.~Berman and D.~C.~Thompson,
  ``Duality Symmetric String and M-Theory,''
  Phys.\ Rept.\  {\bf 566} (2014) 1
  [arXiv:1306.2643 [hep-th]].



\bibitem{Hohm:2013bwa}
  O.~Hohm, D.~L\"ust and B.~Zwiebach,
  ``The Spacetime of Double Field Theory: Review, Remarks, and Outlook,''
  Fortsch.\ Phys.\  {\bf 61} (2013) 926
  [arXiv:1309.2977 [hep-th]].



\bibitem{Baguet:2016prz}
  A.~Baguet, M.~Magro and H.~Samtleben,
  ``Generalized IIB supergravity from exceptional field theory,''
  arXiv:1612.07210 [hep-th].




\bibitem{Hull:1985rc}
  C.~M.~Hull and P.~K.~Townsend,
  ``Finiteness and Conformal Invariance in Nonlinear $\sigma$ Models,''
  Nucl.\ Phys.\ B {\bf 274} (1986) 349.



\bibitem{Tseytlin:1986tt}
  A.~A.~Tseytlin,
  ``Conformal Anomaly in Two-Dimensional Sigma Model on Curved Background and Strings,''
  Phys.\ Lett.\ B {\bf 178} (1986) 34.



\bibitem{Shore:1986hk}
  G.~M.~Shore,
  ``A Local Renormalization Group Equation, Diffeomorphisms, and Conformal Invariance in $\sigma$ Models,''
  Nucl.\ Phys.\ B {\bf 286} (1987) 349.



\bibitem{Tseytlin:1986ws}
  A.~A.~Tseytlin,
  ``$\sigma$ Model Weyl Invariance Conditions and String Equations of Motion,''
  Nucl.\ Phys.\ B {\bf 294} (1987) 383.



\bibitem{Lee:2015qza}
  K.~Lee,
  ``Towards Weakly Constrained Double Field Theory,''
  Nucl.\ Phys.\ B {\bf 909} (2016) 429
  [arXiv:1509.06973 [hep-th]].



\bibitem{Ma:2016vgq}
  C.~T.~Ma and F.~Pezzella,
  ``Supergravity with Doubled Spacetime Structure,''
  arXiv:1611.03690 [hep-th].



\bibitem{Park:2015bza}
  J.~H.~Park, S.~J.~Rey, W.~Rim and Y.~Sakatani,
  ``O(D, D) covariant Noether currents and global charges in double field theory,''
  JHEP {\bf 1511} (2015) 131
  [arXiv:1507.07545 [hep-th]].



\bibitem{Kwak:2010ew}
  S.~K.~Kwak,
  ``Invariances and Equations of Motion in Double Field Theory,''
  JHEP {\bf 1010} (2010) 047
  [arXiv:1008.2746 [hep-th]].



\bibitem{Callan:1985ia}
  C.~G.~Callan, Jr., E.~J.~Martinec, M.~J.~Perry and D.~Friedan,
  ``Strings in Background Fields,''
  Nucl.\ Phys.\ B {\bf 262} (1985) 593.



\bibitem{Tseytlin:1990nb}
  A.~A.~Tseytlin,
  ``Duality Symmetric Formulation of String World Sheet Dynamics,''
  Phys.\ Lett.\ B {\bf 242} (1990) 163.



\bibitem{Tseytlin:1990va}
  A.~A.~Tseytlin,
  ``Duality symmetric closed string theory and interacting chiral scalars,''
  Nucl.\ Phys.\ B {\bf 350} (1991) 395.



\bibitem{Berman:2007xn}
  D.~S.~Berman, N.~B.~Copland and D.~C.~Thompson,
  ``Background Field Equations for the Duality Symmetric String,''
  Nucl.\ Phys.\ B {\bf 791} (2008) 175
  [arXiv:0708.2267 [hep-th]].



\bibitem{Berman:2007yf}
  D.~S.~Berman and D.~C.~Thompson,
  ``Duality Symmetric Strings, Dilatons and O(d,d) Effective Actions,''
  Phys.\ Lett.\ B {\bf 662} (2008) 279
  [arXiv:0712.1121 [hep-th]].



\bibitem{Copland:2011yh}
  N.~B.~Copland,
  ``Connecting T-duality invariant theories,''
  Nucl.\ Phys.\ B {\bf 854} (2012) 575
  [arXiv:1106.1888 [hep-th]].



\bibitem{Copland:2011wx}
  N.~B.~Copland,
  ``A Double Sigma Model for Double Field Theory,''
  JHEP {\bf 1204} (2012) 044
  [arXiv:1111.1828 [hep-th]].



\bibitem{Blair:2015eba}
  C.~D.~A.~Blair,
  ``Conserved Currents of Double Field Theory,''
  JHEP {\bf 1604} (2016) 180
  [arXiv:1507.07541 [hep-th]].



\bibitem{Wald:1993nt}
  R.~M.~Wald,
  ``Black hole entropy is the Noether charge,''
  Phys.\ Rev.\ D {\bf 48} (1993) no.8,  R3427
  [gr-qc/9307038].



\bibitem{Iyer:1994ys}
  V.~Iyer and R.~M.~Wald,
  ``Some properties of Noether charge and a proposal for dynamical black hole entropy,''
  Phys.\ Rev.\ D {\bf 50} (1994) 846
  [gr-qc/9403028].



\bibitem{Iyer:1995kg}
  V.~Iyer and R.~M.~Wald,
  ``A Comparison of Noether charge and Euclidean methods for computing the entropy of stationary black holes,''
  Phys.\ Rev.\ D {\bf 52} (1995) 4430
  [gr-qc/9503052].



\bibitem{Arvanitakis:2016zes}
  A.~S.~Arvanitakis and C.~D.~A.~Blair,
  ``Black hole thermodynamics, stringy dualities and double field theory,''
  arXiv:1608.04734 [hep-th].



\bibitem{West:2001as}
  P.~C.~West,
  ``E(11) and M theory,''
  Class.\ Quant.\ Grav.\  {\bf 18} (2001) 4443
  [hep-th/0104081].



\bibitem{West:2003fc}
  P.~C.~West,
  ``E(11), SL(32) and central charges,''
  Phys.\ Lett.\ B {\bf 575} (2003) 333
  [hep-th/0307098].



\bibitem{West:2004st}
  P.~C.~West,
  ``The IIA, IIB and eleven-dimensional theories and their common E(11) origin,''
  Nucl.\ Phys.\ B {\bf 693} (2004) 76
  [hep-th/0402140].



\bibitem{Hull:2007zu}
  C.~M.~Hull,
  ``Generalised Geometry for M-Theory,''
  JHEP {\bf 0707} (2007) 079
  [hep-th/0701203].



\bibitem{Pacheco:2008ps}
  P.~Pires Pacheco and D.~Waldram,
  ``M-theory, exceptional generalised geometry and superpotentials,''
  JHEP {\bf 0809} (2008) 123
  [arXiv:0804.1362 [hep-th]].



\bibitem{Hillmann:2009ci}
  C.~Hillmann,
  ``Generalized E(7(7)) coset dynamics and D=11 supergravity,''
  JHEP {\bf 0903} (2009) 135
  [arXiv:0901.1581 [hep-th]].



\bibitem{Berman:2010is}
  D.~S.~Berman and M.~J.~Perry,
  ``Generalized Geometry and M theory,''
  JHEP {\bf 1106} (2011) 074
  [arXiv:1008.1763 [hep-th]].



\bibitem{Berman:2011pe}
  D.~S.~Berman, H.~Godazgar and M.~J.~Perry,
  ``SO(5,5) duality in M-theory and generalized geometry,''
  Phys.\ Lett.\ B {\bf 700} (2011) 65
  [arXiv:1103.5733 [hep-th]].



\bibitem{Berman:2011cg}
  D.~S.~Berman, H.~Godazgar, M.~Godazgar and M.~J.~Perry,
  ``The Local symmetries of M-theory and their formulation in generalised geometry,''
  JHEP {\bf 1201} (2012) 012
  [arXiv:1110.3930 [hep-th]].



\bibitem{Berman:2011jh}
  D.~S.~Berman, H.~Godazgar, M.~J.~Perry and P.~West,
  ``Duality Invariant Actions and Generalised Geometry,''
  JHEP {\bf 1202} (2012) 108
  [arXiv:1111.0459 [hep-th]].



\bibitem{Berman:2012vc}
  D.~S.~Berman, M.~Cederwall, A.~Kleinschmidt and D.~C.~Thompson,
  ``The gauge structure of generalised diffeomorphisms,''
  JHEP {\bf 1301} (2013) 064
  [arXiv:1208.5884 [hep-th]].



\bibitem{Park:2013gaj}
  J.~H.~Park and Y.~Suh,
  ``U-geometry: SL(5),''
  JHEP {\bf 1304} (2013) 147
   Erratum: [JHEP {\bf 1311} (2013) 210]
  [arXiv:1302.1652 [hep-th]].



\bibitem{Aldazabal:2013mya}
  G.~Aldazabal, M.~Gra\~na, D.~Marqu\'es and J.~A.~Rosabal,
  ``Extended geometry and gauged maximal supergravity,''
  JHEP {\bf 1306} (2013) 046
  [arXiv:1302.5419 [hep-th]].



\bibitem{Hohm:2013pua}
  O.~Hohm and H.~Samtleben,
  ``Exceptional Form of D=11 Supergravity,''
  Phys.\ Rev.\ Lett.\  {\bf 111} (2013) 231601
  [arXiv:1308.1673 [hep-th]].



\bibitem{Hohm:2013vpa}
  O.~Hohm and H.~Samtleben,
  ``Exceptional Field Theory I: $E_{6(6)}$ covariant Form of M-Theory and Type IIB,''
  Phys.\ Rev.\ D {\bf 89} (2014) no.6,  066016
  [arXiv:1312.0614 [hep-th]].



\bibitem{Hohm:2013uia}
  O.~Hohm and H.~Samtleben,
  ``Exceptional field theory. II. E$_{7(7)}$,''
  Phys.\ Rev.\ D {\bf 89} (2014) 066017
  [arXiv:1312.4542 [hep-th]].



\bibitem{Godazgar:2014nqa}
  H.~Godazgar, M.~Godazgar, O.~Hohm, H.~Nicolai and H.~Samtleben,
  ``Supersymmetric E$_{7(7)}$ Exceptional Field Theory,''
  JHEP {\bf 1409} (2014) 044
  [arXiv:1406.3235 [hep-th]].



\bibitem{Hohm:2014fxa}
  O.~Hohm and H.~Samtleben,
  ``Exceptional field theory. III. E$_{8(8)}$,''
  Phys.\ Rev.\ D {\bf 90} (2014) 066002
  [arXiv:1406.3348 [hep-th]].



\bibitem{Musaev:2014lna}
  E.~Musaev and H.~Samtleben,
  ``Fermions and supersymmetry in E$_{6(6)}$ exceptional field theory,''
  JHEP {\bf 1503} (2015) 027
  [arXiv:1412.7286 [hep-th]].



\bibitem{Hohm:2015xna}
  O.~Hohm and Y.~N.~Wang,
  ``Tensor hierarchy and generalized Cartan calculus in SL(3) $\times$ SL(2) exceptional field theory,''
  JHEP {\bf 1504} (2015) 050
  [arXiv:1501.01600 [hep-th]].



\bibitem{Abzalov:2015ega}
  A.~Abzalov, I.~Bakhmatov and E.~T.~Musaev,
  ``Exceptional field theory: $SO(5,5)$,''
  JHEP {\bf 1506} (2015) 088
  [arXiv:1504.01523 [hep-th]].



\bibitem{Musaev:2015ces}
  E.~T.~Musaev,
  ``Exceptional field theory: $SL(5)$,''
  JHEP {\bf 1602} (2016) 012
  [arXiv:1512.02163 [hep-th]].



\bibitem{Berman:2015rcc}
  D.~S.~Berman, C.~D.~A.~Blair, E.~Malek and F.~J.~Rudolph,
  ``An action for F-theory: $\mathrm{SL}(2)\times {{\mathbb{R}}}^{+}$ exceptional field theory,''
  Class.\ Quant.\ Grav.\  {\bf 33} (2016) no.19,  195009
  [arXiv:1512.06115 [hep-th]].



\bibitem{Ciceri:2016dmd}
  F.~Ciceri, A.~Guarino and G.~Inverso,
  ``The exceptional story of massive IIA supergravity,''
  JHEP {\bf 1608} (2016) 154
  [arXiv:1604.08602 [hep-th]].



\bibitem{Baguet:2016jph}
  A.~Baguet and H.~Samtleben,
  ``E$_{8(8)}$ Exceptional Field Theory: Geometry, Fermions and Supersymmetry,''
  JHEP {\bf 1609} (2016) 168
  [arXiv:1607.03119 [hep-th]].



\bibitem{Coimbra:2012af}
  A.~Coimbra, C.~Strickland-Constable and D.~Waldram,
  ``Supergravity as Generalised Geometry II: $E_{d(d)} \times \mathbb{R}^+$ and M theory,''
  JHEP {\bf 1403} (2014) 019
  [arXiv:1212.1586 [hep-th], arXiv:1212.1586].



\bibitem{Aldazabal:2011nj}
  G.~Aldazabal, W.~Baron, D.~Marques and C.~Nunez,
  ``The effective action of Double Field Theory,''
  JHEP {\bf 1111} (2011) 052
   Erratum: [JHEP {\bf 1111} (2011) 109]
  [arXiv:1109.0290 [hep-th]].



\bibitem{Geissbuhler:2011mx}
  D.~Geissbuhler,
  ``Double Field Theory and N=4 Gauged Supergravity,''
  JHEP {\bf 1111} (2011) 116
  [arXiv:1109.4280 [hep-th]].



\bibitem{Grana:2012rr}
  M.~Grana and D.~Marques,
  ``Gauged Double Field Theory,''
  JHEP {\bf 1204} (2012) 020
  [arXiv:1201.2924 [hep-th]].



\bibitem{Hohm:2014qga}
  O.~Hohm and H.~Samtleben,
  ``Consistent Kaluza-Klein Truncations via Exceptional Field Theory,''
  JHEP {\bf 1501} (2015) 131
  [arXiv:1410.8145 [hep-th]].
\end{thebibliography}
\end{document}